\documentclass[twocolumn]{aastex6}

\usepackage{comment}
\usepackage{amsmath,amssymb}
\usepackage{bm}

\newcommand{\kepler}{{\it Kepler}}
\newcommand{\ktwo}{{\it K2}}

\newcommand{\Mch}{M_\mathrm{ch}}
\newcommand{\Mwd}{M_{\rm WD}}

\newcommand{\ulog}{\mathcal{U}_\mathrm{log}}
\newcommand{\ulin}{\mathcal{U}}

\begin{document}


\title{{Discovery of Three Self-lensing Binaries from \kepler}}
\author{Hajime Kawahara\altaffilmark{1,2}} 
\author{Kento Masuda\altaffilmark{3,6}} 
\author{Morgan MacLeod\altaffilmark{4,7}}
\author{David W. Latham\altaffilmark{4}}
\author{Allyson Bieryla\altaffilmark{4}}
\author{Othman Benomar\altaffilmark{5}}

\altaffiltext{1}{Department of Earth and Planetary Science, 
The University of Tokyo, Tokyo 113-0033, Japan}
\altaffiltext{2}{Research Center for the Early Universe, 
School of Science, The University of Tokyo, Tokyo 113-0033, Japan}
\altaffiltext{3}{Department of Astrophysical Sciences, Princeton University, Princeton, NJ 08544, USA} 
\altaffiltext{4}{Harvard-Smithsonian Center for Astrophysics, Cambridge, MA 02138, USA}
\altaffiltext{5}{Center for Space Science, NYUAD Institute, New York University Abu Dhabi, PO Box 129188 Abu Dhabi, UAE}
\altaffiltext{6}{NASA Sagan Fellow}
\altaffiltext{7}{NASA Einstein Fellow}

\email{Electronic address: kawahara@eps.s.u-tokyo.ac.jp}
\slugcomment{Accepted for publication in The Astronomical Journal on Dec 23.}
\shortauthors{Kawahara et al.}
\shorttitle{Self-lensing Binaries from Kepler}

\begin{abstract}
  {We report the discovery of three edge-on binaries with white dwarf companions that gravitationally magnify (instead of eclipsing) the light of their stellar primaries, as revealed by a systematic search for pulses with long periods in the \kepler\ photometry.  We jointly model the self-lensing light curves and radial-velocity orbits to derive the white dwarf masses, all of which are close to 0.6 Solar masses. The orbital periods are long, ranging from 419 to 728 days, and the eccentricities are low, all less than 0.2. These characteristics are reminiscent of the orbits found for many blue stragglers in open clusters and the field, for which stable mass transfer due to Roche-lobe overflow from an evolving primary (now a white dwarf) has been proposed as the formation mechanism.  Because the actual masses for our three white dwarf companions have been accurately determined, these self-lensing systems would provide excellent tests for models of interacting binaries.}

\end{abstract}


\section{Introduction}\label{sec:intro}

Photometry by the {\it Kepler} spacecraft with unprecedented precision uncovered thousands of transiting planets and eclipsing stars. The latter includes a dozen short-period eclipsing binaries with low mass white dwarf (WD) secondaries \citep{2010ApJ...713L.150R, 2010ApJ...715...51V,2011MNRAS.410.1787B,2011ApJ...728..139C,2012ApJ...748..115B,2013ApJ...767..111M,2015ApJ...803...82R,2015ApJ...815...26F}. These compact systems are believed to have undergone mostly stable mass transfer from the WD progenitor to the current primary because they are located near the theoretical orbital period--white dwarf mass relation for the stable mass transfer (see Section \ref{sec:evolution} for more details). 

As the orbital period increases, the gravitational lensing effect by the WD becomes important in modeling the eclipse light curve \citep[e.g.][]{2011MNRAS.410.1787B}. 
{When the lensing magnification surpasses the dimming due to a normal eclipse, the light curves can even exhibit positive pulses, rather than dips, as the WD passes in front of its stellar companion.} 
\citet{2014Sci...344..275K} found periodic pulses in the \kepler\ Object of Interest (KOI)-3278 system and identified it as a self-lensing binary (SLB) of a WD and a main-sequence (MS) G-type star on a 88-day orbit --- as was predicted almost 50 years ago \citep{1969ApJ...156.1013T, 1971AaA....15..251L, 1973AaA....26..215M} and has been studied theoretically by many authors \citep{1995ApJ...441...77G,2001MNRAS.324..547M,2002ApJ...579..430A,2002AaA...394..489B,2003ApJ...584.1042S,2003ApJ...592.1151F,2003ApJ...594..449A,2011MNRAS.410..912R,2016ApJ...820...53H}. Unlike the compact \kepler\ WD--MS eclipsing binaries of suggested stable mass transfer origin, KOI-3278 is considered to be a post common-envelope binary \citep[PCEB;][]{2014AaA...568L...9Z} similar to many other WDs with red-dwarf companions from SDSS \citep[e.g.][and references therein]{2007MNRAS.382.1377R,2010ApJS..190..275F,2010MNRAS.402..620R,2013AJ....146...82R,2014AaA...570A.107R,2016MNRAS.458.3808R}.
The PCEB is an outcome of unstable mass transfer, in which the runaway transfer rate precludes accretion onto the MS star and leads to a shared, common envelope. The final orbit shrinks because of dynamical friction forces between the binary and the envelope \citep{1976IAUS...73...75P}.

The present paper demonstrates a further diversity of the post-interaction binaries, reporting the discovery of three long-period ($1$--$2\,\mathrm{yr}$) SLBs from a systematic pulse survey in the {\it Kepler} light curves. The WD mass and orbital period derived from the light curve and follow-up radial velocity (RV) observations suggest that the three SLBs likely experienced binary interactions --- however, 
{their wide orbits suggest that the interactions did not lead to common-envelope evolution as proposed for KOI-3278 and other PCEBs.}

The rest of the paper is organized as follows. In Section \ref{sec:search}, we describe the procedure to detect repeating pulses in the {\it Kepler} light curves and to confirm them with follow-up RV observations. In Section \ref{sec:joint}, we model the observed pulses and RVs to derive the lens mass and its orbit assuming that the lens is a WD. In Section \ref{sec:evolution}, we discuss the physical properties of the SLBs in the context of the binary evolution model. Section \ref{sec:summary} summarizes and concludes the paper.

\section{Search and Validation of Self-lensing Pulses in the \kepler\ Light Curves}\label{sec:search}

We searched for periodic pulse signals via visual inspection combined with an auto-correlation analysis of the light curve (Section \ref{ssec:acf}) because the widely-used box-least square algorithm is not optimized to identify a small number ($\lesssim3$) of repeating signals. This procedure allowed us to detect six candidate objects listed in Table \ref{tab:candidates}. Note that the pulse period of KIC 8145411 was not uniquely determined due to a data gap between the two detected pulses. For each candidate, we performed the  ephemeris-matching test (Section \ref{ssec:ephemeris}) and centroid-shift analysis (Appendix \ref{ssec:centroid}). These tests identified KIC 8622134 as a false positive.

\begin{figure*}[htbp]
\begin{center}
\includegraphics[bb=0 0 891 405,width=\linewidth]{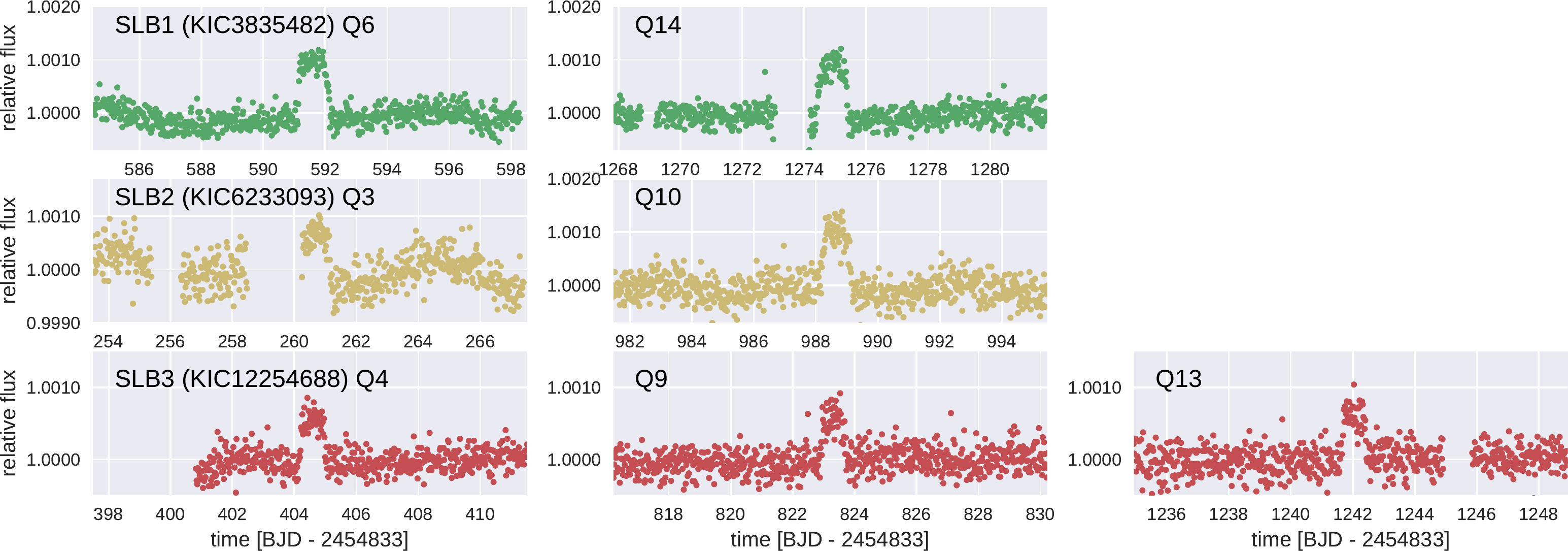}
\caption{The observed pulses of the SLBs. The light curves are PDCSAP flux of the {\it Kepler} long-cadence data.\label{fig:allp}}
\end{center}
\end{figure*}

We performed follow-up RV observations of the remaining five systems (Section \ref{ssec:vetting_rv}) and detected RV variations consistent with the pulse light curve for three of them, which we confirmed to be genuine self-lensing binaries (SLBs 1--3; Figure \ref{fig:allp}). KIC 6522276 showed no RV variations and is also likely a false positive. Further follow-up observations are required to determine the status of KIC 8145411.

\begin{table}[!tbh]
\begin{center}
\caption{SLB Candidates from the Pulse Search\label{tab:candidates}}
\begin{tabular}{cccccccc}
\hline\hline
          & KIC & \# of Pulses & Period (day) & $Kp$ & \\
\hline
SLB1  &  3835482 &2 & 683.27 & 13.2 & \\ 
SLB2  &  6233093 &2 & 727.98 & 13.9 & \\ 
SLB3  & 12254688 &3 & 418.72 & 13.1 & \\ 
  &  8145411 &2 & 455.84 & 14.6 \\ 
  & 			&3 & 911.67 & 14.6 & \\ 
 FP &  6522276 &2 & 768.55 & 14.7 & \\ 
 FP &  8622134 &3 & 374.55 & 15.7 & \\ 
\hline
\end{tabular}
\end{center}
{Note --- Due to a data gap at the midpoint of the two detected pulses, the orbital period of KIC 8145411 is not uniquely determined from the light curve. The follow-up RV observations did not solve the degeneracy either (Figure \ref{fig:SLB4RV}). {$Kp$ indicates \kepler\ magnitude.}}
\end{table}

\subsection{Pulse Identification}\label{ssec:acf}

{We analyzed the long cadence ($29.4\,\mathrm{min}$) PDCSAP light curves for all targets and all quarters from the primary \kepler\ mission. No \ktwo\ light curves were searched.} Each light curve was divided into two parts A ($f_A$) and B ($f_B$) at the midpoint and smoothed by the median filter with the width of three bins (1.5 hr).
Then we searched for features common in light curves A and B (i.e., repeating pulses) by computing their windowed cross-correlation function:
\begin{eqnarray}
  C(\tau,p) = \int_0^T dt \, w(t-\tau) f_A(t) \, f_B(t-p) \quad \text{for} \quad 0 \le p \le T,\nonumber\\
\end{eqnarray}
where $w(t)$ is the box car window with width $L$, and $T$ is the time length of the data. We chose $L=1.5\,\mathrm{days}$ to match a typical {pulse} duration of long-period objects. The typical number of data points in each segment is 35,000 and $C(\tau,\theta)$ has $\sim 10^9$ data points for each star in the {\it Kepler} input catalog (KIC). Using a graphic processing unit (NVIDIA Geforce Titan X), we computed $C(\tau,p)$ for all the KIC stars ($N \approx 2 \times 10^5$). If the maximum value of the cross-correlation function at $(\tau, p)=(T_0, P)$ exceeded the 5$\sigma$ level, we visually inspected the light curve around $t_1 = T_0$ and $t_2 = T_0 + P$. About half of the targets ($N \sim 10^5$) satisfied the $5\sigma$ criterion.

We detected repeating pulses in the light curves of six KIC stars (Table \ref{tab:candidates}). We detected a pair of pulses for KIC 3835482, KIC 6233093, KIC 8145411, and KIC 6522276. The cross-correlation analysis detected two pulses for KIC 12254688 and KIC 8622134, and we identified the third ones by visual inspection. 

\subsection{Ephemeris Matching} \label{ssec:ephemeris}

\begin{figure*}[htbp]
\begin{center}
\includegraphics[bb=0 0 1080 346,width=0.9\linewidth]{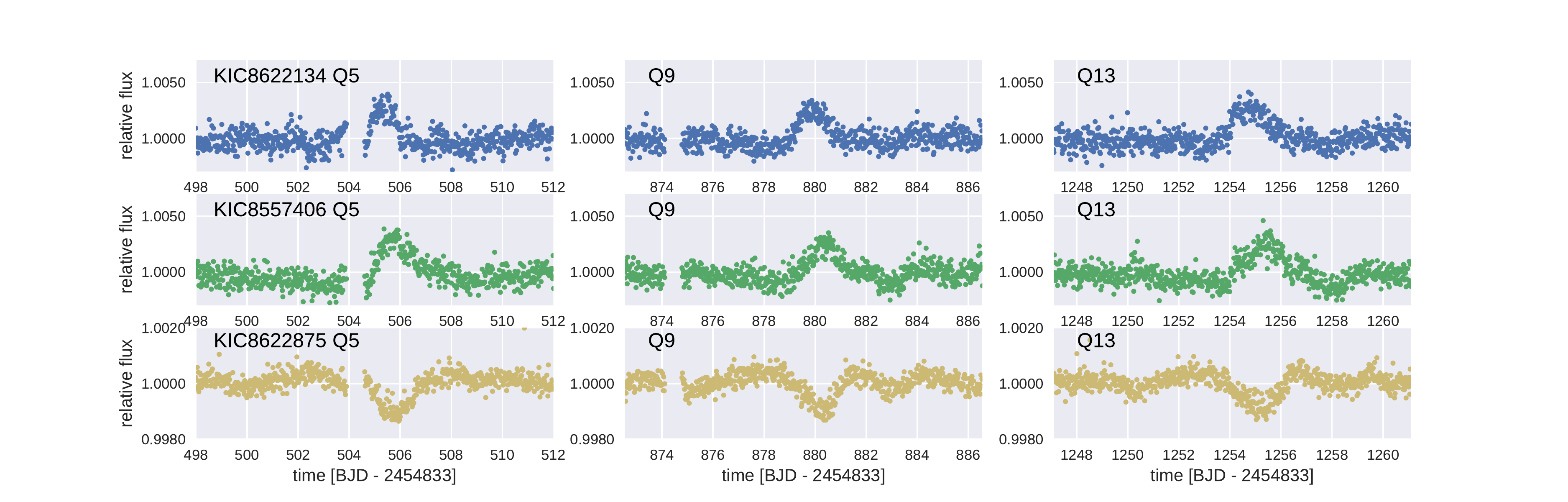}
\caption{Pulses of KIC 8622134 compared to those of the neighbor stars KIC 8557406 and KIC 8622875 (KOI-5551). The latter two were identified by ephemeris matching. \label{fig:emmatch}}
\end{center}
\end{figure*}

We found that KIC 8557406, located close to KIC 8622134, has pulses at the same timings as those of KIC 8622134 (Figure \ref{fig:emmatch}). This indicates that the pulses of KIC 8622134 are actually caused by contamination. This is also supported by the centroid offset during the pulses (Appendix \ref{ssec:centroid}). Thus we exclude this target from the SLB candidates. These features might be attributed to the cross-talk between different CCD pixels. For example, KIC 10989166 and KIC 10989274 are known to have the inverted signals from the eclipsing binary KIC 9851142.
 
We performed similar tests for the remaining five candidates and found that no known KOI, {\it Kepler} eclipsing binary \citep{2016AJ....151...68K}, or long-period transiting planet candidate \citep{2014AJ....148...28S,2015ApJ...815..127W,2016AJ....152..206F,2016ApJ...822....2U} has an ephemeris that matches those of the candidates.
{We also inspected the long cadence PDCSAP light curves of the relatively bright neighbor stars of the candidates (KIC 3835487 for KIC 3835482, KIC 6233055 for KIC 6233093, and KIC 6522279/6522242/6522288 for KIC 6522276, all located within 10 pixels). Again we found no suspicious signals at the timings of the pulses.}

\subsection{Radial Velocity (RV) Observations}\label{ssec:vetting_rv}

We obtained high-resolution spectra to measure RVs of KIC 3835482, KIC 6233093, KIC 6522276, KIC 8145411, and KIC 12254688 with the Tillinghast Reflector Echelle Spectrograph (TRES) on the 1.5~m telescope at the Fred Lawrence Whipple Observatory (FLWO) in Arizona.  
The observed RV values are listed in Table \ref{tab:rvdata}. As 
will be shown in Section \ref{ssec:joint},
KIC 3835482, KIC 6233093, and KIC 12254688 show clear velocity variations consistent with the pulse light curves. 
Thus we conclude that these three targets are genuine SLBs and denote them as SLBs 1--3. Their light curves around the detected pulses are shown in Figure \ref{fig:allp}.

\begin{table*}[!tbh]
\begin{center}
\caption{Radial Velocities of the SLB Candidates\label{tab:rvdata}}
\begin{tabular}{ccccc}
\hline\hline
& KIC & Time (BJD$_\mathrm{TDB}$) & RV (km/s) & Error (km/s)\\
\hline
SLB1 & 3835482 &  2457673.6811 &  $-15.88$  & $0.04$\\
 &  &  2457837.0037  &  $-7.35$ & $0.10$\\
 &  &  2457852.9552  &   $-5.99$ & $0.06$\\
 &  &  2457887.9108  &  $-3.14$ &  $0.05$\\
 &  &  2457900.9508  &   $-2.14$ &  $0.04$\\
 &  &  2457933.8526  &   $0.15$ &  $0.09$\\
 &  &  2457993.7576  &   $2.18$  &  $0.05$\\
 &  &  2458003.6854  &   $2.16$  &  $0.05$\\
 &  &  2458019.7216  &   $2.07$ &  $0.06$\\
 &  &  2458041.6358	&	$1.42$ &  $0.04$\\
 &  &  2458053.6572	&	$0.97$ &  $0.05$\\
 &  &  2458069.6052	&	$0$ &  $0.05$\\
 &  &  2458083.6118  &  $-0.87$  &  $0.07$\\
 \\
SLB2 & 6233093 &  2457909.9111 & $13.64$ & $0.06$\\
 & & 2457936.8489 & $12.15$ & $0.10$\\
 & & 2457994.8278 & $7.13$ & $0.12$\\
 & & 2458008.7673 & $5.71$ & $0.08$\\
 & & 2458020.7421 & $4.49$ & $0.07$\\
 & & 2458040.6404 & $2.36$ & $0.09$\\
 & & 2458063.5964 & $0.32$ & $0.11$\\
 & & 2458069.6519 & $0$ & $0.09$\\
 \\
SLB3 & 12254688 & 2457901.7927 & $-12.08$ & $0.09$\\
 &  & 2457933.8829 & $-8.05$ & $0.10$\\
 &  & 2457993.7827 & $-1.83$ & $0.09$\\
 &  & 2457993.7310 & $-1.23$ & $0.09$\\
 & & 2458019.6670 & $0$ & $0.09$\\
 & & 2458040.6821 & $1.08$ & $0.12$\\
 & & 2458052.6794 & $1.61$ & $0.08$\\
 & & 2458067.5907 & $1.85$ & $0.10$\\
 & & 2458080.5934 & $1.64$ &  $0.08$\\
 \\
& 6522276 & 2457909.8662 & $0.17$ & $0.10$\\
& & 2457936.8072 & $0$ & $0.15$\\
 & & 2458020.7883 & $0.02$ & $0.15$\\
 \\
 & 8145411 & 2457900.8368 & $1.13$ & $0.11$\\
&  & 2457935.7773 & $-0.18$ & $0.14$\\
&  & 2458007.6608 & $0$ & $0.14$\\
& & 2458021.7223 & $0.98$ & $0.11$\\
 & & 2458067.6248 & $4.4$ & $0.09$\\
\hline
\end{tabular}
\end{center}
{Note --- The RV values are multi-order velocities relative to the observation with the highest S/N ratio (with zero RV value) used as the template in the cross-correlation analysis.  The quoted errors are the internal precision based on the RMS scatter in the relative velocity.}
\end{table*}

\begin{figure*}[htbp]
\begin{center}
\includegraphics[bb=0 0 605 236,width=0.9\linewidth]{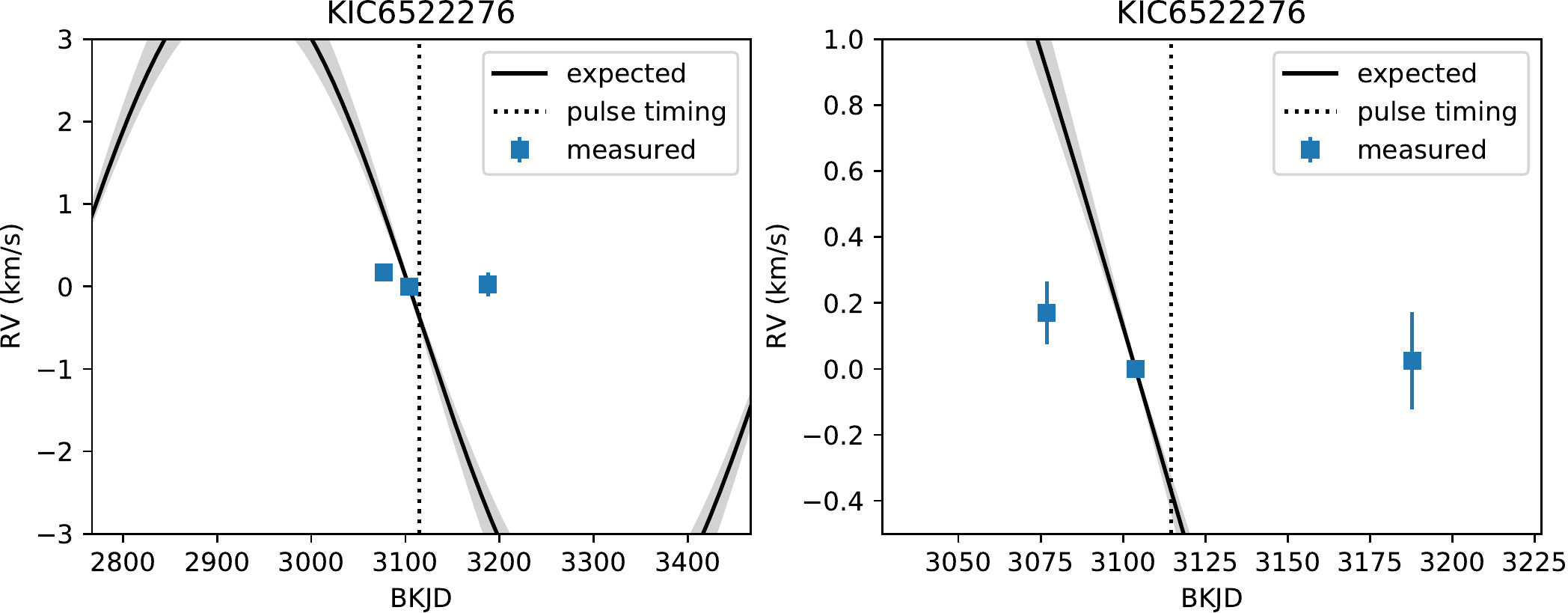}
\caption{
Comparison of the observed RVs of KIC 6522276 (blue squares) to the model expected from the light curve. The shaded region around the solid line shows the $68\%$ uncertainty of the prediction from the light-curve model, and the systemic velocity in the model is fixed to match the second data point. The error bars on the first and third points indicate the error relative to the second measurement. The right panel is an enlarged version of the left one. The vertical dashed line indicates the phase of the pulse. \label{fig:SLC3RV}}
\end{center}
\end{figure*}

\begin{figure*}[htbp]
\begin{center}
\includegraphics[bb=0 0 891 220, width=0.9\linewidth]{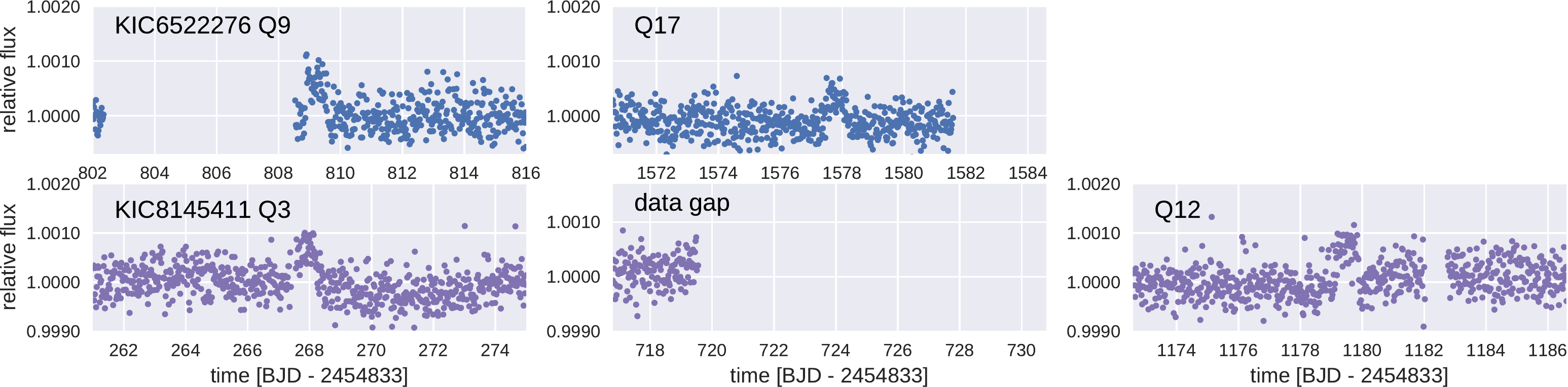}
\caption{Pulses detected in the light curves of KIC 6522276 (top) and KIC 8145411 (bottom).\label{fig:SLC3pulse}}
\end{center}
\end{figure*}

Unlike SLBs 1--3, KIC 6522276 did not exhibit significant RV variation more than $\sim0.2$~km/s. The null detection is inconsistent with the presence of a stellar-mass companion (Figure \ref{fig:SLC3RV}). Thus we conclude that KIC 6522276 is a false positive. The slight difference in the heights of two observed pulses (Figure \ref{fig:SLC3pulse}, top) also supports this notion.

\begin{figure*}[htbp]
\begin{center}
\includegraphics[bb=0 0 419 348, width=0.43\linewidth]{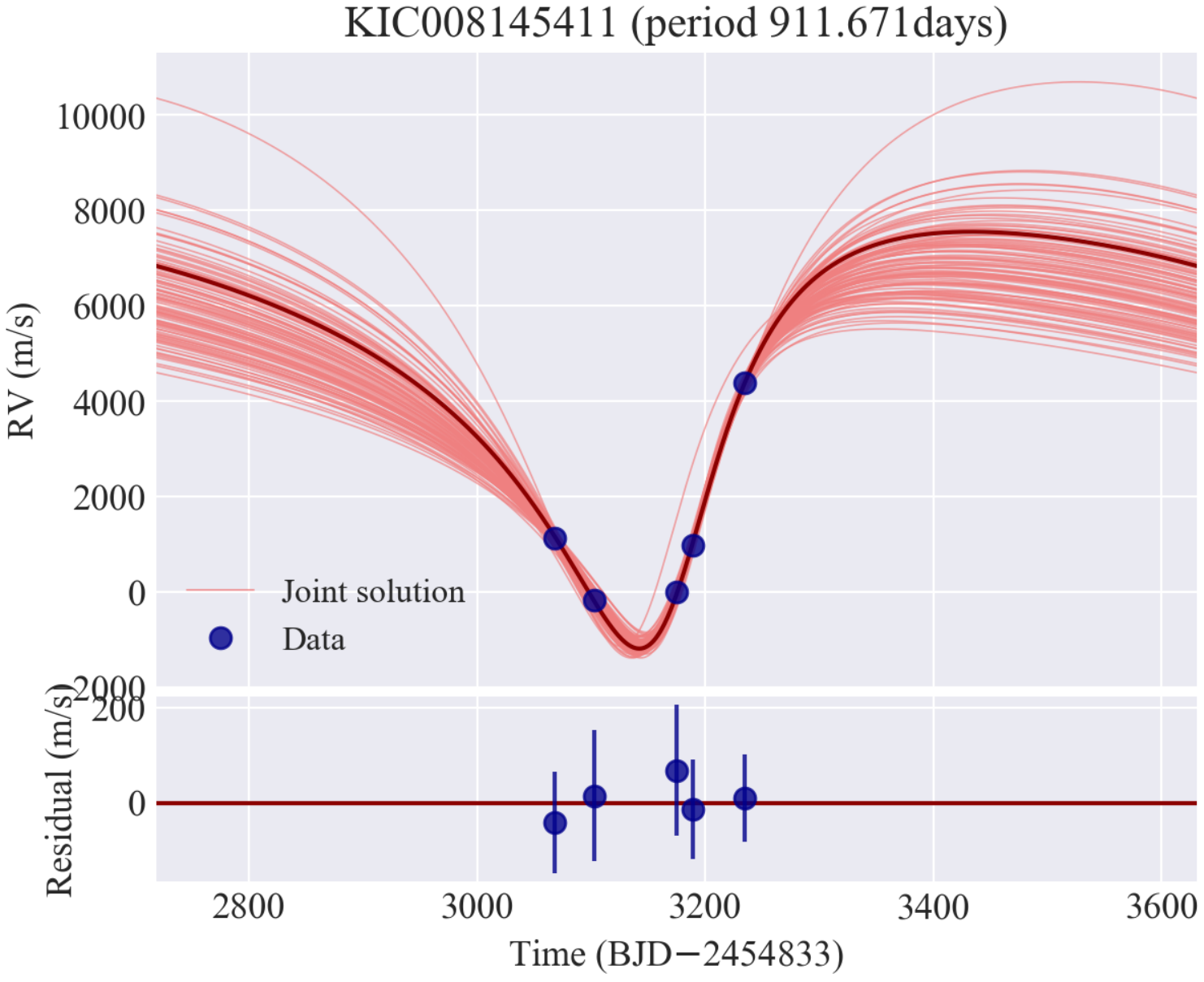}\hspace{0.7cm}
\includegraphics[bb=0 0 425 348, width=0.43\linewidth]{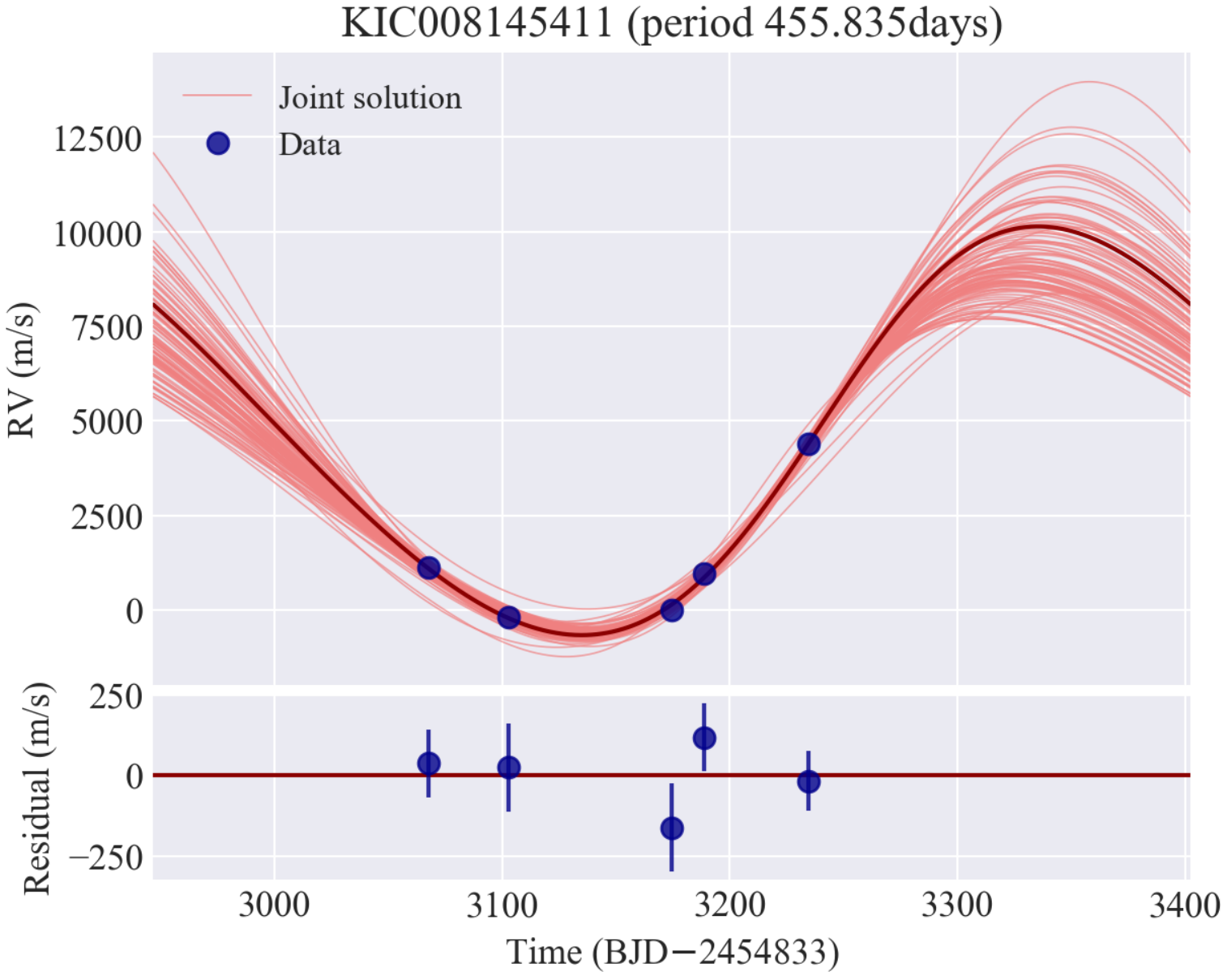}
\caption{
Observed RVs of KIC 8145811 (blue circles) and random $100$ posterior models (red solid lines) with two different orbital periods allowed from the light curve (left: $P=911.671\,\mathrm{days}$, right: $P=455.835\,\mathrm{days}$). The thick red solid line corresponds to the maximum likelihood model, for which residuals in the bottom panels are calculated. The models are from the joint fit to the \kepler\ light curves and RVs; see Section \ref{sec:joint} for details of the model.
\label{fig:SLB4RV}}
\end{center}
\end{figure*}

KIC 8145411, on the other hand, did show an RV variation (Figure \ref{fig:SLB4RV}), but the observation could be consistent with either of the models with two different orbital periods ($P=911.7\,\mathrm{days}$ and $P=455.8\,\mathrm{days}$) allowed from the light curves (Figure \ref{fig:SLC3pulse}, bottom), if the eccentricity is adjusted accordingly. Given this uncertainty in the orbital solution, as well as a low S/N of the pulse signal, we consider this target to be an unconfirmed candidate until more RV data are acquired.

\subsection{Planetary Companion in SLB 3?}

SLB 3 is identified as KOI-2384 because of a possible periodic transit-like signal with $P=62.6823\,$days. Although the {\it Kepler} team classified this object KOI-2384.01 to be a false positive based on the centroid offset, the significance was rather marginal, and our own centroid analysis described in Appendix \ref{ssec:centroid} did not find conclusive evidence either. We analyzed transit timing variations (TTVs) of {KOI-2384.01} and found no significant variation larger than about $0.01\,$days. Since this limit is much smaller than the TTV amplitude expected from a stellar-mass companion of SLB 3 at $P=419\,$days, we conclude that {KOI-2384.01} is a false positive{, as originally thought. The origin of the signal is unclear, although the small nominal value for the third-light contamination \citep[CROWDSAP metric of $1.0000$;][]{2016ksci.rept....9T} may imply that it is either due to cross talk or contamination from an unresolved third star.}

\section{Physical Properties of the SLBs}\label{sec:joint}

We derived the parameters of the three SLBs by jointly modeling their phase-folded pulse light curves and RVs (Figure \ref{fig:slc_fits}), assuming that the compact companions are WDs. The results are summarized in Table \ref{tab:slbs}. Unlike KOI-3278 \citep{2014Sci...344..275K}, we did not detect secondary eclipses of the WDs, although the inferred eccentricity is small and the impact parameter during the secondary eclipse is likely smaller than unity. Thus we also examined the consistency between the upper limit on the WD luminosity (i.e. lower limit on the cooling age) and other inferred properties of the system (Section \ref{ssec:lc_luminosity}). We found no discrepancy between the inferred age and system dimensions, at least within current uncertainties.

\subsection{Characterization of the Primary Stars} \label{ssec:tres}

For primary properties, we adopt the stellar parameters from KIC DR25 \citep{2016arXiv160904128M} as the prior information (Table \ref{tab:primary}). 
These values are broadly consistent with the atmospheric parameters based on the TRES spectra, although we did not use the latter in the analysis because of their large uncertainties due to low S/N.
We attempted to detect asteroseismic oscillations {in the power spectra of the light curves, but did not find conclusive evidence of pulsations (Appendix \ref{ssec:seismology}).}

We will later argue that our SLBs have likely been formed via mass transfer to the stellar primaries. This means that the unknown quantity of mass gain by the primary might affect the mass determination through comparison to normal, single-star evolution isochrones. In the following, we simply adopt the values based on the single-star model because we do not expect a deviation more than the current precision. The possible deviation, if any, will come into sharp focus once the primary masses are measured dynamically by combining RV data with the self-lensing light curves.

\subsection{Modeling of the {\it Kepler} Light Curves and RV Data}\label{ssec:joint}

\begin{figure*}[htbp]
\begin{center}
\includegraphics[width=0.48\linewidth,bb=0 0 547 425]{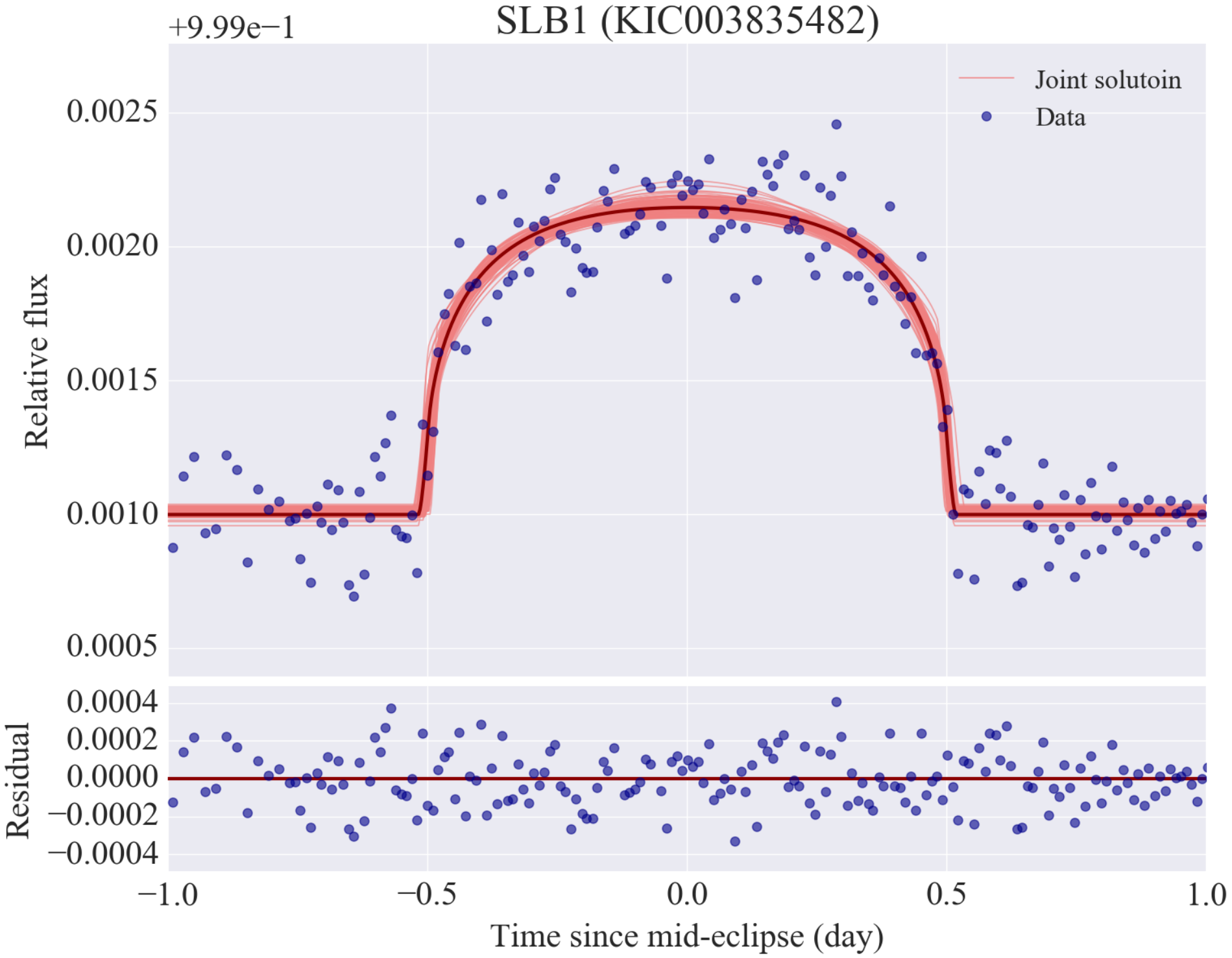}\hspace{0.5cm}
\includegraphics[width=0.46\linewidth,bb=0 0 444 348]{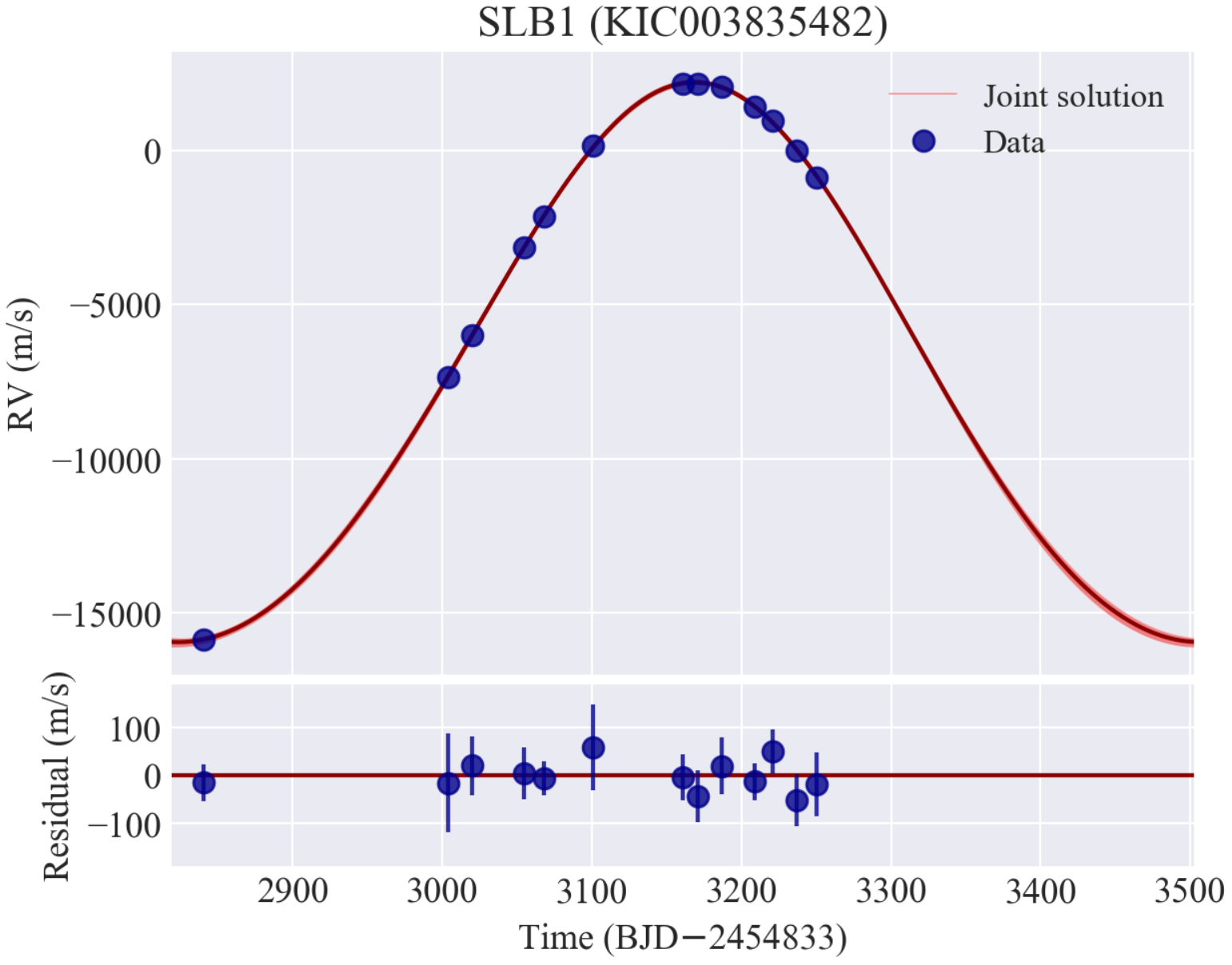}
\includegraphics[width=0.48\linewidth,bb=0 0 547 425]{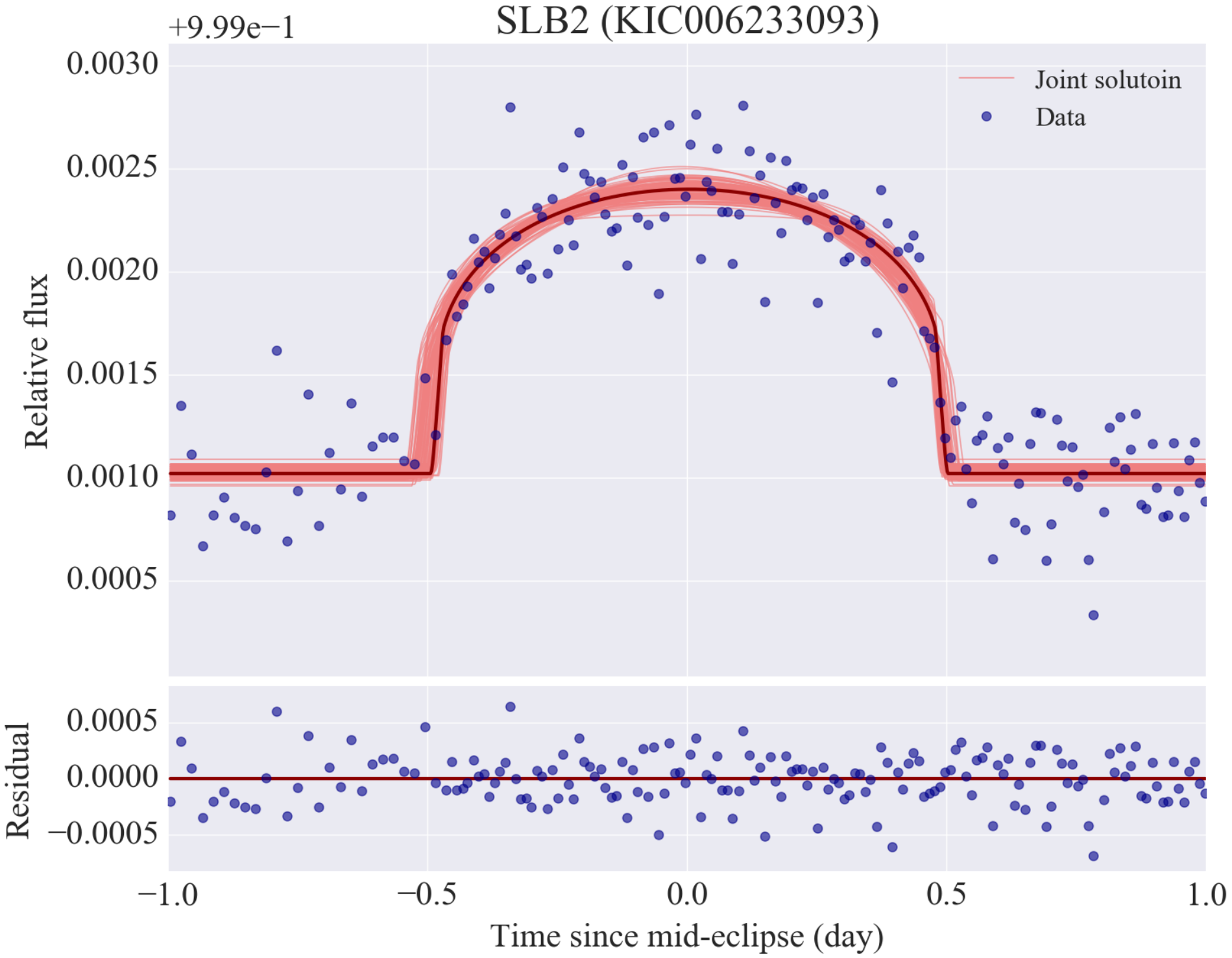}\hspace{0.5cm}
\includegraphics[width=0.46\linewidth,bb=0 0 444 348]{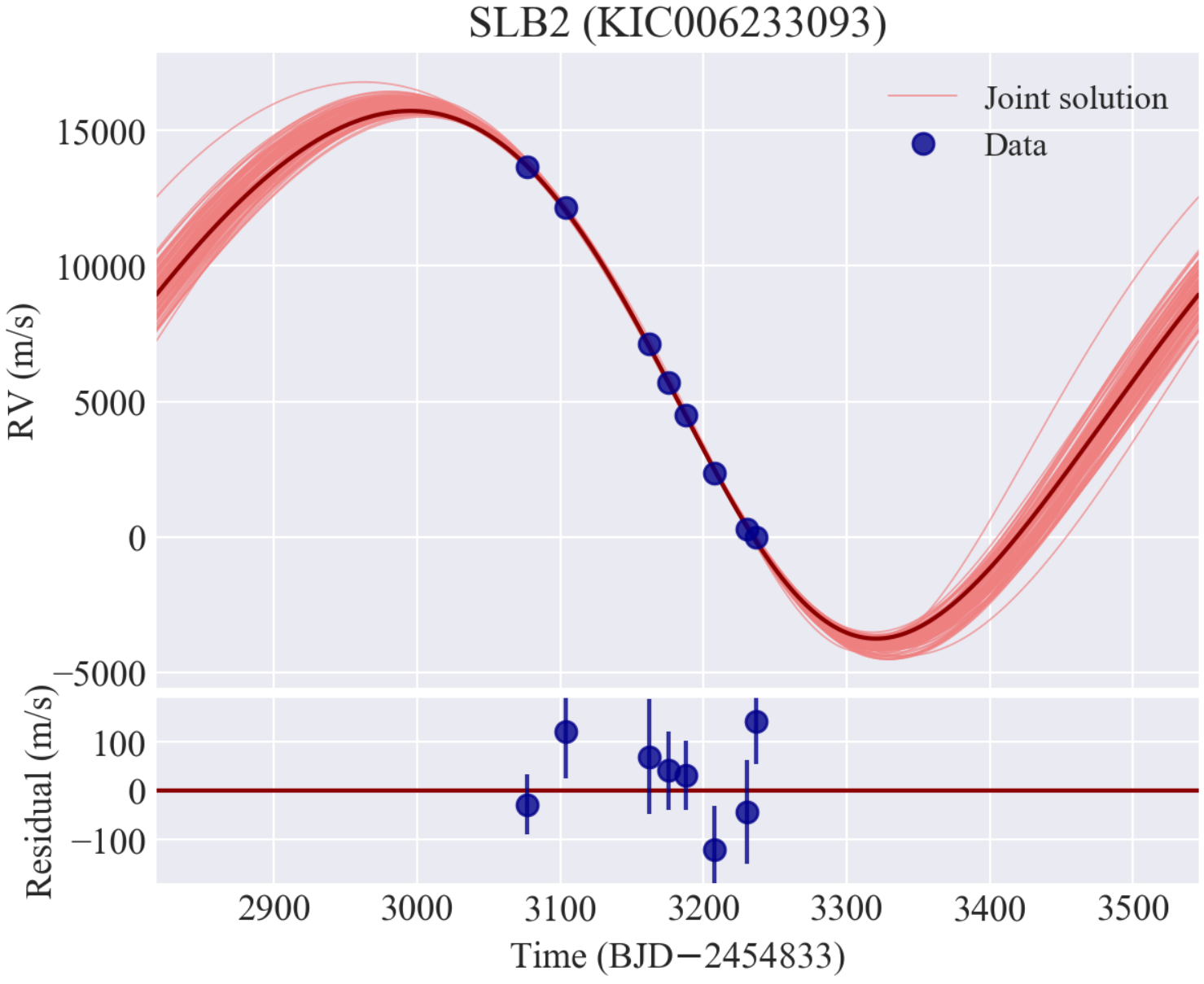}
\includegraphics[width=0.48\linewidth,bb=0 0 547 425]{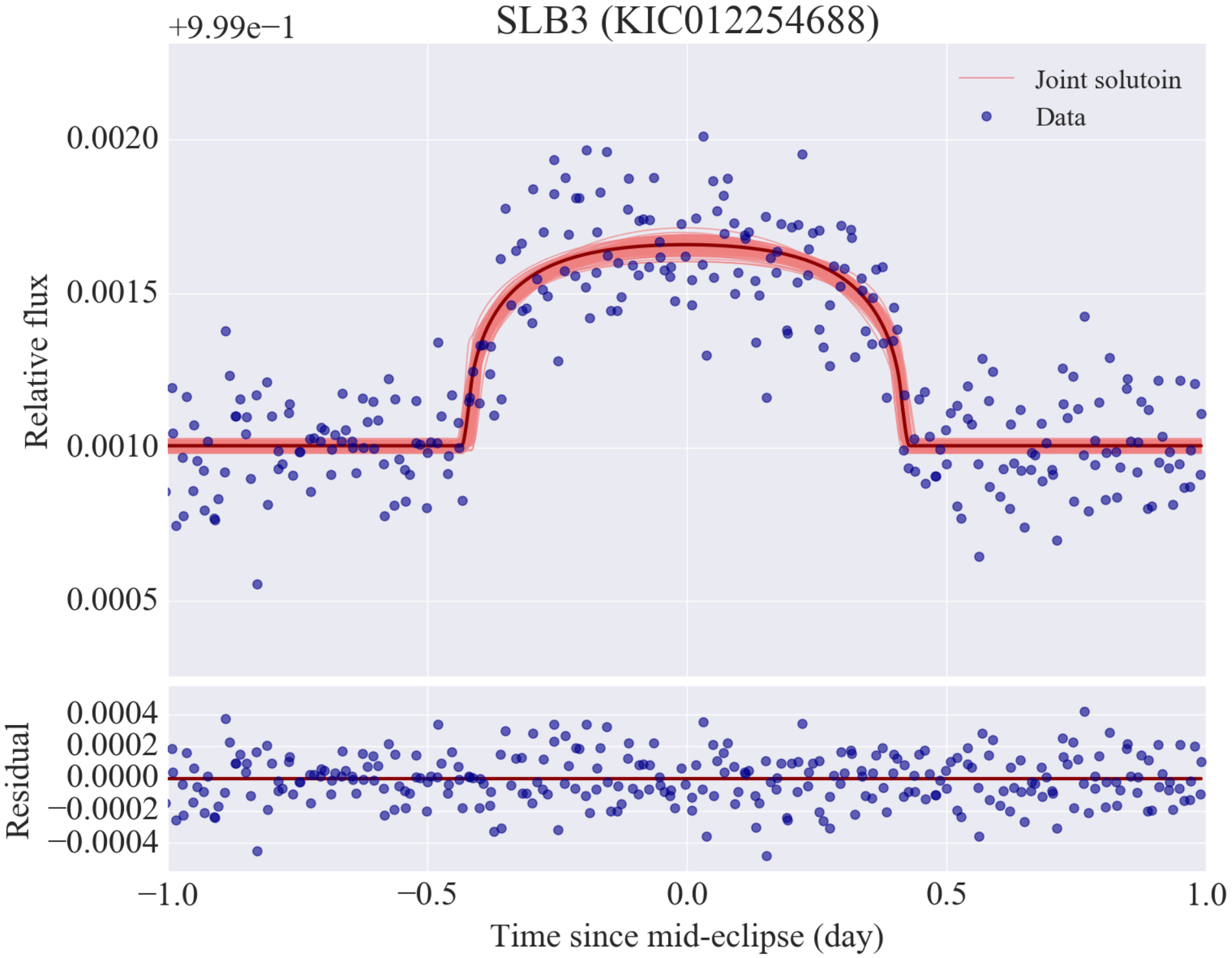}\hspace{0.5cm}
\includegraphics[width=0.46\linewidth,bb=0 0 444 348]{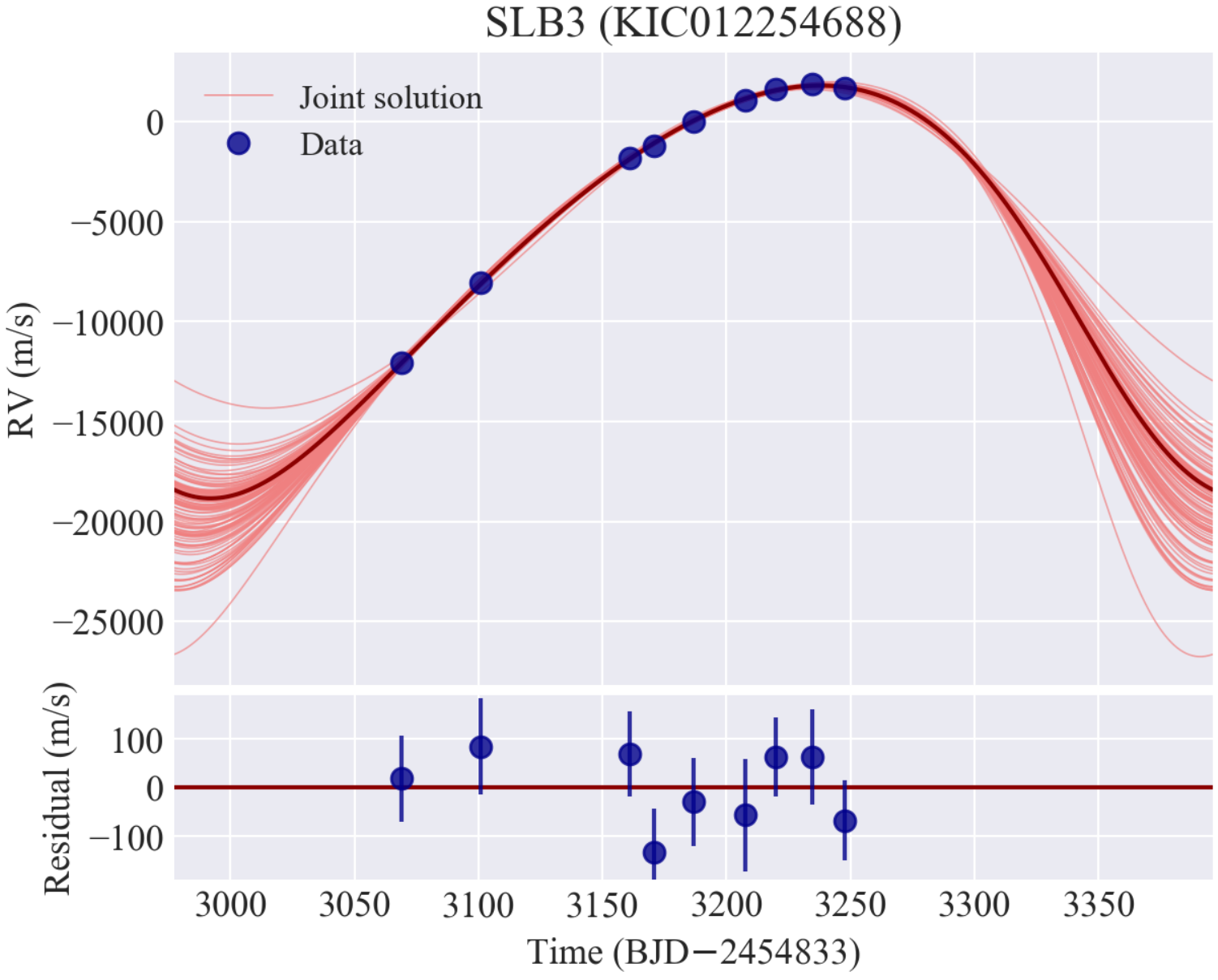}
\caption{Joint modeling of detrended and phase-folded {\it Kepler} light curves (left column) and RVs (right column) of three SLBs. In both columns, the blue dots show the data points and the red solid lines are random $100$ posterior models from the joint analysis. The thick red solid line corresponds to the maximum likelihood model, for which residuals in the bottom panels are calculated.\label{fig:slc_fits}}
\end{center}
\end{figure*}

The parameters of the SLB systems (Table \ref{tab:slbs}) were derived via a joint analysis of the phase-folded {\it Kepler} light curves (Figure \ref{fig:slc_fits} left) and the RV time series (Table \ref{tab:rvdata}, Figure \ref{fig:slc_fits} right). For the light curves, we used Simple Aperture Photometry (SAP) fluxes downloaded from the NASA Exoplanet Archive. Only the long cadence data were available for the SLBs. We iteratively detrended the light curves around the pulses using a second-order polynomial \citep{2015ApJ...805...28M}, and stacked them around their central times to obtain a single pulse light curve.

We used a combination of the non-linear least-squares fitting \citep{2009ASPC..411..251M} and the Markov Chain Monte Carlo (MCMC) sampling \citep{2013PASP..125..306F} to derive the posterior probability distribution for the model parameters. Summary statistics for the marginalized posteriors are found in Table \ref{tab:slbs}, and the posterior models are compared with the data in Figure \ref{fig:slc_fits}. The likelihood of the model was computed by
\begin{align}
	\notag
	\mathcal{L}=&\frac{1}{\sqrt{2\pi (\sigma_{{\rm SAP},i}^2+\tilde\sigma_{\rm LC}^2)}}
	\prod_i \exp\left[-\frac{(f_{{\rm data},i}-f_{{\rm model},i})^2}{2(\sigma_{{\rm SAP},i}^2+\tilde\sigma_{\rm LC}^2)}\right]\\
	\label{eq:likelihood}
	&\times
	\frac{1}{\sqrt{2\pi (\sigma_{{\rm RV},j}^2+\tilde\sigma_{\rm RV}^2)}} \prod_j \exp\left[-\frac{(v_{{\rm data},j}-v_{{\rm model},j})^2}{2(\sigma_{{\rm RV},j}^2+\tilde\sigma_{\rm RV}^2)}\right].
\end{align}
Here $f$ and $v$ with the subscripts data/model stand for the flux and RV data/model, respectively; $\sigma_{\rm SAP}$ is the error in the SAP flux; $\sigma_{\rm RV}$ is the internal RV error given in Table \ref{tab:rvdata}; $\tilde\sigma_{\rm LC}$ and $\tilde\sigma_{\rm RV}$ take into account any additional scatter in the data that is not included in the model. Both were optimized and marginalized over to derive conservative constraints on the other physical parameters. We adopted independent, non-informative (uniform or log-uniform) priors for all the fitted parameters in Table \ref{tab:slbs} except for $M_\star$ and $\rho_\star$; for these two parameters, the joint posterior distribution provided by the KIC DR25 was adopted.

The light curve around the pulse was modeled as
\begin{align}
	f_{\rm model}=c_{\rm pulse}(1+f_{\rm pulse}-f_{\rm eclipse}),
\end{align}
where $c_{\rm pulse}$ is the normalization of the light curve. The modulation was modeled as a superposition of the brightening due to lensing ($f_{\rm pulse}$) and the usual dimming due to an eclipse ($f_{\rm eclipse}$). This approximation is valid as long as the Einstein radius of the WD, 
\begin{align}
	\notag
	R_{\rm E} &= \sqrt{\frac{4GM_{\rm WD}}{c^2} \frac{a(1-e^2)}{1+e\sin\omega}}\\
	\notag
	&= 4.7\,R_\oplus \left( \frac{M_{\rm WD}}{M_\odot} \right)^{1/2}
	\left( \frac{M_{\rm WD} + M_\star}{M_\odot} \right)^{1/3} \left(\frac{P}{\mathrm{1 yr}}\right)^{2/3}\\ 
	\label{eq:re}
	&\times \sqrt{\frac{1-e^2}{1+e\sin\omega}},
\end{align}
is not too different from the physical WD radius and $R_{\rm E}\ll R_\star$ \citep{2016ApJ...820...53H}. The eclipse part was computed using {\tt PyTransit} package \citep{2015MNRAS.450.3233P} from the limb-darkening coefficients $q_1$ and $q_2$ \citep{2013MNRAS.435.2152K}, mass $M_\star$ and mean density $\rho_\star$ of the primary star, eclipse impact parameter $b$, physical WD radius $R_{\rm WD}$, time of inferior conjunction $t_{\rm c}$, orbital period $P$, orbital eccentricity $e$, and argument of periastron $\omega$ referred to the sky plane. Following \citet{2003ApJ...594..449A}, the pulse part was approximated as an inverted transit: we adopt $R_{\rm E}$ as the radius of the eclipsing object to compute the usual transit light curve, scale the depth by a factor of two, and flip it around the baseline. In computing the eclipse part, the WD radius was derived from its mass using the Eggleton mass--radius relation (Appendix \ref{sec:mrrelation}).

{We neglect the possible third-light contamination considering that the CROWDSAP metrics \citep{2016ksci.rept....9T}, which gives the ratio of target flux relative to flux from all sources within the photometric aperture, are smaller than about $1\%$ during any pulse event for all three SLBs. This simplification does not affect our current results, which involve larger uncertainties in the physical parameters (see also Section \ref{sssec:norv}).}

The RV variation of the primary star was modeled assuming a Keplerian orbit, whose zero point $\gamma_{\rm RV}$ was fitted to match the observed RVs. In our analysis this is not the systemic velocity because RVs given in Table \ref{tab:rvdata} are values relative to the observation with the highest S/N ratio.

\begin{table*}[tbh]
\begin{center}
\caption{Physical Properties of Primary Stars from KIC DR25\label{tab:primary}}
\begin{tabular}{ccccccc}
\hline\hline
& KIC & $T_{\rm eff}$ (K)  & $\log g$ (cgs) & $M_\star$ ($M_\odot$) & $R_\star$ ($R_\odot$) & $\rho_\star$ (g\,$\mathrm{cm}^{-3}$)\\
\hline
SLB1 & 03835482 & $6017^{+190}_{-232}$  & $4.271^{+0.158}_{-0.193}$ & $1.1060\pm0.1510$ 
	& $1.275^{+0.367}_{-0.267}$ & $0.7515^{+0.5908}_{-0.3816}$\\
SLB2 & 06233093 & $5937^{+196}_{-178}$  & $3.931^{+0.443}_{-0.148}$ & $1.0960^{+0.1650}_{-0.2010}$
	& $1.876^{+0.451}_{-0.902}$ & $0.2338^{+1.016}_{-0.1029}$\\
SLB3 &12254688 & $7058^{+191}_{-233}$ & $3.648^{+0.328}_{-0.082}$ & $1.5620^{+0.2420}_{-0.3230}$
	& $3.101^{+0.398}_{-1.113}$ & $0.07373^{+0.1801}_{-0.01963}$\\
\hline
\end{tabular}
\end{center}
\end{table*}

\begin{table*}[!tbh]
\begin{center}
\caption{System Parameters of SLBs from RVs and \kepler\ Light Curves\label{tab:slbs}}
\begin{tabular}{lcccccc}
\hline\hline
Parameter & SLB1  & SLB2 & SLB3 && Prior\tablenotemark{a}\\
& 03835482 & 06233093 & 12254688 && \\
\hline
\multicolumn{2}{l}{\it (Fixed Parameters)}\\
Orbital period $P$ (day)\tablenotemark{b}	& {$683.267(7)$} & {$727.98(1)$} & {$418.715(6)$} && $\cdots$ \\
Eclipse epoch $t_0$ ($\mathrm{BJD_{TDB}}$)\tablenotemark{b} & {$2455424.628(5)$} & {$2455093.68(1)$} & {$2455237.584(7)$} && $\cdots$\\ 
Primary effective temperature $T_\star$ (K) & $6017$ & $5937$ & $7058$ && $\cdots$\\
\\
\multicolumn{2}{l}{\it (Fitted Parameters)}\\
Primary mass $M_\star$ ($M_\odot$) & $1.2^{+0.3}_{-0.1}$ & $1.2^{+0.2}_{-0.1}$ & $1.5\pm0.1$ && KIC DR25\\
Primary mean density $\rho_\star$ (g\,$\mathrm{cm}^{-3}$) & $0.24^{+0.01}_{-0.02}$ & $0.24\pm0.02$ & $0.19^{+0.03}_{-0.02}$ && KIC DR25\\
White dwarf mass $M_{\rm WD}$ ($M_\odot$) & $0.53^{+0.08}_{-0.03}$ & $0.62^{+0.05}_{-0.04}$ & $0.62^{+0.09}_{-0.06}$ && $\ulog(0.01, 1.454)$\\
Limb-darkening coefficient $q_1$ & $0.7\pm0.2$ & $0.5^{+0.3}_{-0.2}$ & $0.7^{+0.2}_{-0.3}$ && $\ulin(0, 1)$\\
Limb-darkening coefficient $q_2$ & $0.2^{+0.2}_{-0.1}$ & $0.4^{+0.3}_{-0.2}$ & $0.3\pm0.2$ && $\ulin(0, 1)$\\
Center of the phased eclipse $t_{\rm c}-t_0$ (day)  & $0.001^{+0.004}_{-0.003}$ & $0.002^{+0.006}_{-0.009}$ & $-0.001\pm0.005$ && $\ulin(-1, 1)$\\
Eclipse impact parameter $b$ & $0.3\pm0.1$ & $0.1\pm0.1$ & $0.09^{+0.09}_{-0.07}$ && $\ulin(0, 1)$\\
Pulse normalization $c_{\rm pulse}$ & $1.00000(1)$ & $1.00003(3) $  & $1.00001(1)$ && $\ulin(0.9995, 1.0005)$\\
Light-curve jitter $\sigma_{\rm LC}$ ($10^{-5}$) & $3^{+3}_{-2}$ & $11^{+3}_{-4}$ & $11\pm1$ && $\ulog(10^{-6}, 10^{-3})$\\
RV jitter $\sigma_{\rm RV}$ ($\mathrm{m\,s^{-1}}$)& $5^{+12}_{-4}$ & $24^{+79}_{-21}$ & $12^{+48}_{-10}$ && $\ulog(1, 10^3)$\\
RV zero point $\gamma_{\rm RV}$ ($\mathrm{m\,s^{-1}}$) & $-7.43\pm0.02$ & $6.7\pm0.3$ & $-7.7^{+0.4}_{-0.5}$ && $\ulin(-1.5\times10^4, 1.5\times10^4)$\\
$\sqrt{e}\cos\omega$ & $0.244\pm0.005$ & $-0.23^{+0.07}_{-0.06}$  & $-0.28\pm0.04$ && $\ulin(-1, 1)$\tablenotemark{c}\\
$\sqrt{e}\sin\omega$ & $0.04\pm0.02$ & $0.21^{+0.04}_{-0.06}$  & $0.34\pm0.03$ && $\ulin(-1, 1)$\tablenotemark{c}\\
\\
\multicolumn{2}{l}{\it (Derived Parameters)}\\
Primary radius $R_\star$ ($R_\odot$) & $1.88^{+0.21}_{-0.09}$ & $1.9\pm0.1$ & $2.2
^{+0.2}_{-0.1}$ && $\cdots$\\
White dwarf radius $R_{\rm WD}$ ($R_\odot$) & $0.0133^{(+5)}_{(-10)}$ & $0.0122^{(+8)}_{(-11)}$ & $0.011(1)$ && $\cdots$\\
Total mass $M_\star+M_{\rm WD}$ ($M_\odot$) & $1.7^{+0.4}_{-0.2}$ & $1.8\pm0.2$ & $2.2\pm0.2$ && $\cdots$\\
Orbital eccentricity $e$ & $0.061\pm0.001$ & $0.10\pm0.02$ & $0.20\pm0.04$ && $\cdots$\\
Argument of pericenter $\omega$ (deg) & $8\pm4$ & $137^{+13}_{-15}$ & $129\pm2$ && $\cdots$\\
RV amplitude $K_\star$ ($\mathrm{km\,s^{-1}}$) & $9.09\pm0.02$ & $10.0^{+0.3}_{-0.2}$ & $10.8^{+1.0}_{-0.8}$ && $\cdots$\\
\hline
\end{tabular}
\end{center}
{Note --- The quoted values show the median of the marginal posterior distribution and $68.3\%$ interval around it. Values in the parentheses show the uncertainty in the last digit. }
\tablenotetext{a}{Priors adopted in the MCMC sampling. $\ulin(x,y)$ and $\ulog(x,y)$ denote the uniform and log-uniform distributions between $x$ and $y$, respectively. For $M_\star$ and $\rho_\star$, we adopt the joint posterior distribution for these two parameters from KIC DR25 as the prior.}
\tablenotetext{b}{Best linear ephemerides from the iterative phase folding described in Section \ref{ssec:joint}. {The uncertainties are $68.3\%$ intervals of the marginal posteriors from the MCMC fitting.}}
\tablenotetext{c}{The condition $(\sqrt{e}\cos\omega)^2+(\sqrt{e}\sin\omega)^2<1$ was also imposed.}
\end{table*}

\subsubsection{Light-curve Modeling without RV Data}\label{sssec:norv}

{To test the reliability of our light-curve model, we also fit the {\it Kepler} light curves of SLBs 1--3 without RV data. This is useful as a check because any discrepancy with the RV result, if present, points to unmodeled systematics such as significant third-light contamination.} Here we omitted the second part of the likelihood in Eqn.~\ref{eq:likelihood}, but included the light curves around the secondary-eclipse phase assuming $e=0$, which turned out to be a reasonable approximation even with the RV data. This analysis allows us to place rough upper limits on the WD luminosity based on the absence of the secondary eclipse.\footnote{Strictly speaking, non-zero eccentricity also changes the phase of the secondary eclipse. This difference is ignored here because we did not find the secondary eclipses in other phases of the light curve either.}

The detrending of the secondary light curves was performed only with a second-order polynomial, since we did not identify a significant secondary eclipse in any of the systems. Here the light curve was modeled as
\begin{align}
	f_{\rm model}=\begin{cases}
	c_{\rm pulse}(1+f_{\rm pulse}-f_{\rm eclipse})
	&\quad\text{(pulse)}\\c_{\rm secondary}(1-f_{\rm secondary}) &\quad\text{(secondary)}\end{cases},
\end{align}
{where $f_{\rm secondary}$ was computed from the WD radius and effective temperatures of both objects ($T_\star$ and $T_{\rm WD}$), by convolving the Planck function with the response function of {\it Kepler}.}  The results of this analysis are summarized in Table \ref{tab:params_lconly}.

\begin{table*}[!tbh]
\begin{center}
\caption{System Parameters of SLBs from \kepler\ Light Curves Alone \label{tab:params_lconly}}
\begin{tabular}{lcccccc}
\hline\hline
Parameter & SLB1  & SLB2 & SLB3 && Prior\tablenotemark{a}\\
& 03835482 & 06233093 & 12254688 && \\
\hline
\multicolumn{2}{l}{\it (Fixed Parameters)}\\
Orbital period $P$ (day)\tablenotemark{b}	& {$683.267(7)$} & {$727.98(1)$} & {$418.715(6)$} && $\cdots$ \\
Eclipse epoch $t_0$ ($\mathrm{BJD_{TDB}}$)\tablenotemark{b} & {$2455424.628(5)$} & {$2455093.68(1)$} & {$2455237.584(7)$} && $\cdots$\\ 
Primary effective temperature $T_\star$ (K) & $6017$ & $5937$ & $7058$ && $\cdots$\\
\\
\multicolumn{2}{l}{\it (Fitted Parameters)}\\
Primary mass $M_\star$ ($M_\odot$) & $1.3\pm0.2$ & $1.1^{+0.2}_{-0.1}$ & $1.4\pm0.2$ && KIC DR25\\
Primary mean density $\rho_\star$ (g\,$\mathrm{cm}^{-3}$) & $0.26^{+0.03}_{-0.12}$ & $0.27^{+0.05}_{-0.09}$ & $0.20^{+0.10}_{-0.08}$ && KIC DR25\\
White dwarf mass $M_{\rm WD}$ ($M_\odot$) & $0.52^{+0.33}_{-0.04}$ & $0.53^{+0.21}_{-0.05}$ & $0.5^{+0.3}_{-0.1}$ && $\ulog(0.01, 1.454)$\\
White dwarf temperature $T_{\rm WD}$ (K) & $<1\times10^4$  & $<9\times10^3$ & $<1\times10^4$ && $\ulog(1000, 50000)$\\
Limb-darkening coefficient $q_1$ & $0.7\pm0.2$ & $0.6\pm0.3$ & $0.7\pm0.2$ && $\ulin(0, 1)$\\
Limb-darkening coefficient $q_2$ & $0.2^{+0.2}_{-0.1}$ & $0.4^{+0.3}_{-0.2}$ & $0.2^{+0.2}_{-0.1}$ && $\ulin(0, 1)$\\
Center of the phased eclipse $t_{\rm c}-t_0$ (day)  & $0.001\pm0.004$ & $0.000^{+0.006}_{-0.009}$ & $-0.001^{+0.005}_{-0.004}$ && $\ulin(-1, 1)$\\
Eclipse impact parameter $b$ & $0.3^{+0.3}_{-0.2}$ & $0.3^{+0.3}_{-0.2}$ & $0.5^{+0.2}_{-0.3}$ && $\ulin(0, 1)$\\
Pulse normalization $c_{\rm pulse}$ & $1.00000(1)$ & $1.00003(3) $  & $1.00000(1)$ && $\ulin(0.9995, 1.0005)$\\
Secondary normalization $c_{\rm secondary}$ & $1.00000(1)$ & $1.00000(2)$ & $1.000002(9)$ && $\ulin(0.9995, 1.0005)$\\
Light-curve jitter $\sigma_{\rm LC}$ ($10^{-5}$) & $8.5^{+0.9}_{-1.0}$ & $12\pm2$ & $8.2\pm0.7$ && $\ulog(10^{-6}, 10^{-3})$\\
\\
\multicolumn{2}{l}{\it (Derived Parameters)}\\
Primary radius $R_\star$ ($R_\odot$) & $1.9^{+0.5}_{-0.1}$ & $1.8^{+0.4}_{-0.1}$ & $2.1^{+0.5}_{-0.3}$ && $\cdots$\\
Secondary radius $R_{\rm WD}$ ($R_\odot$) & $0.0135^{(+5)}_{(-40)}$ & $0.0134^{(+7)}_{(-26)}$ & $0.013^{(+2)}_{(-3)}$ && $\cdots$\\
Total mass $M_\star+M_{\rm WD}$ ($M_\odot$) & $1.9^{+0.5}_{-0.2}$ & $1.7^{+0.3}_{-0.2}$ & $2.0^{+0.5}_{-0.3}$ && $\cdots$\\
RV amplitude $K_\star$ ($\mathrm{km\,s^{-1}}$) & $8.7^{+3.2}_{-0.6}$ & $9.1^{+2.1}_{-0.8}$ & $10^{+3}_{-2}$ && $\cdots$\\
\hline
\end{tabular}
\end{center}
{Note --- The quoted values show the median of the marginal posterior distribution and $68.3\%$ interval around it except for the effective temperature of the white dwarf $T_{\rm WD}$, for which we show the $99\%$ percentile of the marginal posterior distribution. Values in the parentheses show the uncertainty in the last digit. }
\tablenotetext{a}{Priors adopted in the MCMC sampling. $\ulin(x,y)$ and $\ulog(x,y)$ denote the uniform and log-uniform distributions between $x$ and $y$, respectively. For $M_\star$ and $\rho_\star$, we adopt the joint posterior distribution for these two parameters from KIC DR25 as the prior.}
\tablenotetext{b}{Best linear ephemerides from the iterative phase folding described in Section \ref{ssec:joint}. {The uncertainties are $68.3\%$ intervals of the marginal posteriors from the MCMC fitting.}}
\end{table*}

\subsection{Constraints on Luminosity and Cooling Age}\label{ssec:lc_luminosity}

The null detection of the secondary eclipse in the above analysis places upper limits on the WD temperature $T_{\rm WD}$. As we noted above, the result is based on the assumption of $e=0$, which may not be exactly the case for our SLBs. Nevertheless, these limits can still be useful for a consistency check, because the impact parameter during the secondary eclipse is likely smaller than unity (based on the values of $b$, $e$, and $\omega$) for all three SLBs and because the constraint is not so sensitive to the assumed phase of the secondary eclipse in the case of null detection.

\begin{figure*}[htbp]
\begin{center}
\includegraphics[width=0.45\linewidth,bb=0 0 574 457]{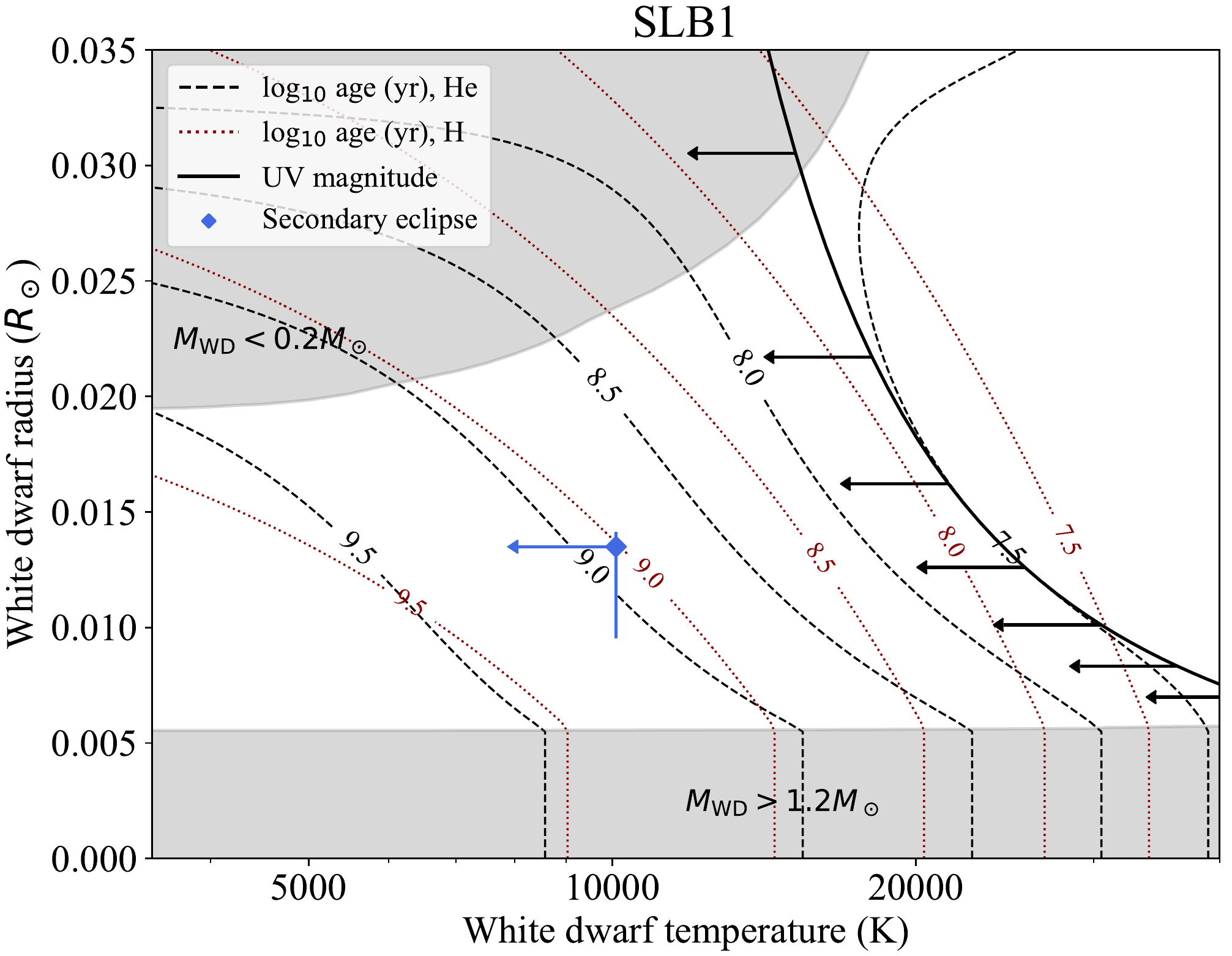}\hspace{0.5cm}
\includegraphics[width=0.45\linewidth,bb=0 0 574 457]{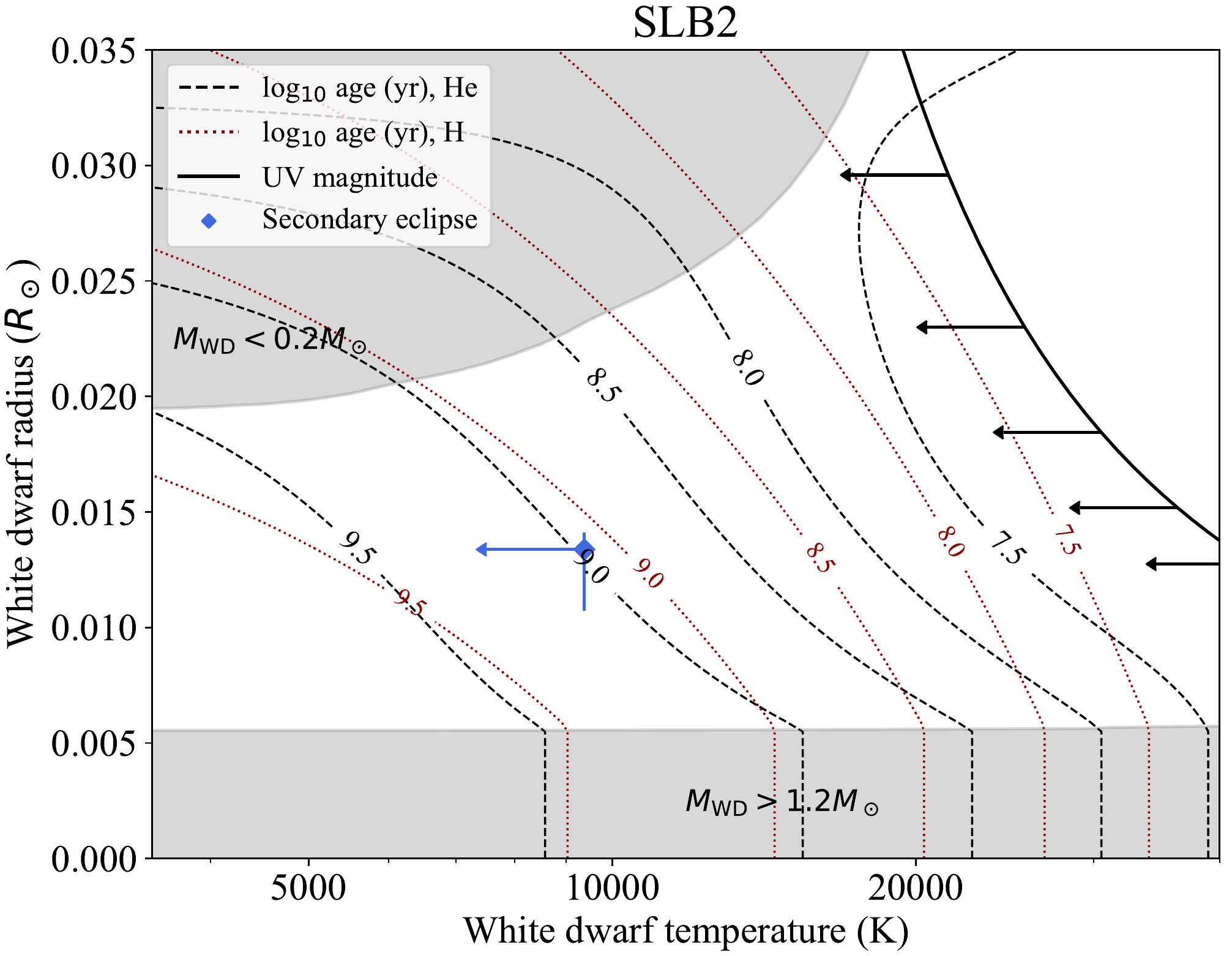}
\includegraphics[width=0.45\linewidth,bb=0 0 574 457]{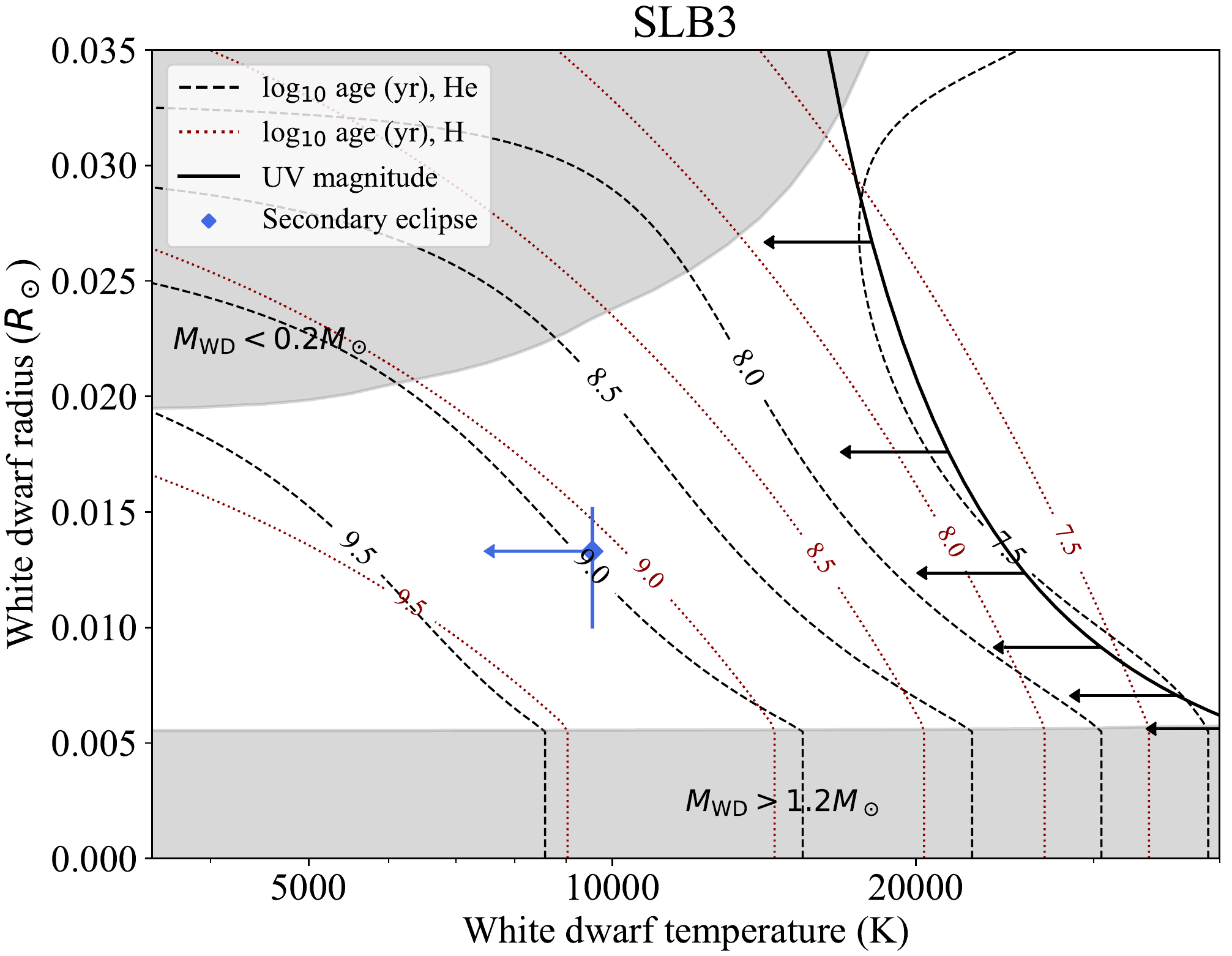}
\caption{Constraints on the $R_{\rm WD}$--$T_{\rm WD}$ plane from the secondary eclipse light curves (blue diamonds) and the broad-band photometry (black lines). Contours of the theoretical cooling age are also shown: black dashed lines are for the pure-He atmosphere, and red dotted lines are for the pure-H atmosphere (see Appendix \ref{sec:mrrelation} for more details).\label{fig:cooling}}
\end{center}
\end{figure*}

Figure \ref{fig:cooling} compares the resulting constraints on $R_{\rm WD}$ ($1\sigma$ credible interval) and $T_{\rm WD}$ ($3\sigma$ upper limits) with theoretical cooling curves of WDs with hydrogen and helium atmospheres (see Appendix \ref{sec:mrrelation}). We find that the non-detections of secondary eclipses are naturally explained if the WD age is more than $\sim$Gyr, with little dependence on the assumed atmospheric composition. These values do not conflict with the dimensions of the primaries (Table \ref{tab:slbs}). Thus the lack of secondary eclipses is compatible with our interpretation of these systems.

We also checked the archival data of the broadband photometry for these candidates (Section \ref{sssec:broad}). These data provide additional constraints on the WD luminosity as shown with black solid lines in Figure \ref{fig:cooling}. The constraints on the age are weaker than those from secondary eclipses typically by $1$--$2$ orders of magnitude, but they are independent from the assumption of $e=0$ and thus are more conservative.

\subsubsection{Analysis of Broadband Spectra}\label{sssec:broad}

We searched for UV excess by modeling the spectral energy distribution from GALEX (FUV and NUV), SDSS ($g$, $r$, $i$, and $z$), 2MASS ($J$, $H$, and $K_s$), and WISE (3.35$\mu$m, 4.6$\mu$m, and 11.6 $\mu$m). We used the virtual observatory SED analyzer \citep{2008AaA...492..277B} with the NextGen model \citep{1999ApJ...512..377H} adopting the values of $d$ and $A_{\rm V}$ from the KIC DR25. Both FUV ($0.15 \mu$m) and NUV ($0.23 \mu$m) are available for SLB 3, while the only NUV is available for the others. The best-fit temperatures and radii are consistent with the values from DR25 within 5\%. We did not find any apparent UV excess for all the SLBs in the GALEX band. Given this non-detection, along with uncertainties in $d$ and $A_{\rm V}$, we decided to use the magnitude of the FUV for SLB 3 and NUV for the rest to place upper limits on the WD luminosity. Assuming the Planck function for the emissivity, we constrain $R_{\rm WD}$ and $T_{\rm WD}$ as
\begin{eqnarray}
  \label{eq:TRconst}
  R_\mathrm{WD}^2 B_\lambda (T_\mathrm{WD}) \le \frac{R_\star^2 F_\lambda}{\pi d^2} ,
\end{eqnarray}
where $B_\lambda (T_\mathrm{WD})$ and $F_\lambda$ are the Planck function and the unreddened flux at the FUV (for SLB 3) or NUV (for the rest) band. We adopted the values of $R_\star$ in Table \ref{tab:primary}.

For SLB 2, we find an IR excess in the WISE 11.6 $\mu$m band. The excess is an order of magnitude larger than the SED model in the flux unit and may imply the presence of dust. 

\section{Binary Evolution History of SLBs}\label{sec:evolution}

\begin{figure*}[htbp]
\begin{center}
\includegraphics[width=\linewidth,bb=0 0 650 432]{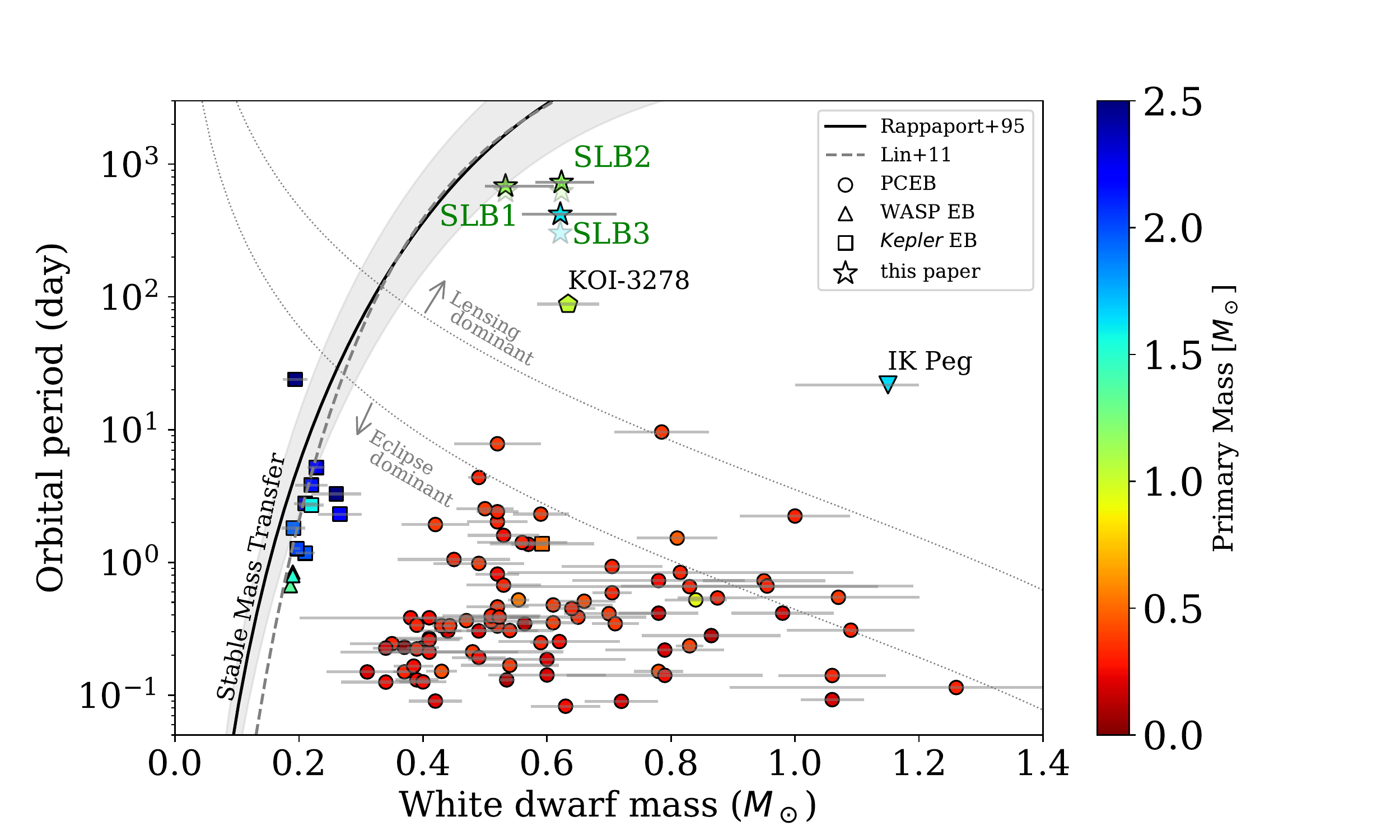}
\caption{Selected WD binaries with stellar companions in the $P$--$M_{\rm WD}$ plane. The star symbols indicate three SLBs reported in the present paper; the pentagon symbol indicates KOI-3278 \citep{2014Sci...344..275K}; the square and triangle symbols are the eclipsing (but not self-lensing) systems from {\it Kepler} \citep{2011MNRAS.410.1787B,2011ApJ...728..139C,2012ApJ...748..115B,2013ApJ...767..111M,2015ApJ...803...82R,2015ApJ...815...26F,2015ApJ...806..155M} and WASP \citep{2011MNRAS.418.1156M,2013Natur.498..463M}; the circles are post common envelope binaries (PCEB) found by SDSS and/or the eclipsing PCEBs \citep{2012MNRAS.419..806R,2013AaA...549A..95Z}; and the inverted triangle is IK Peg \citep{1993PASP..105..841L}. The color indicates the mass of the stellar primary. 
For SLBs, the periods corresponding to the separation at periastron, $P(1-e)^{3/2}$, are also shown with transparent symbols.
The solid line and the shaded region show the theoretical relationship for the stable mass transfer and its upper and lower limits (factors of $2.4$ from the middle curve) from \citet{1995MNRAS.273..731R}, while the dashed line shows the same relation modified for the shorter-period systems \citep{2011ApJ...732...70L}. The gray dotted curves denote the boundaries where the self-lensing pulse is twice larger than eclipse dip (``lensing dominant" line) and where the opposite is the case (``eclipse dominant"), both computed for $M_\star=1\,M_\odot$.\label{fig:pm}}
\end{center}
\end{figure*}

\subsection{Binary Interaction in the Formation of WD Binaries}

Sufficiently close binaries, with separations less than $10^3 R_\odot$, interact when their stellar components evolve and expand. 
We briefly review the  binary interaction pathways thought to shape the formation of WD--MS binaries \citep{2014apa..book.....G}. 
Following its formation and the main sequence evolution of its components, the more massive star in the binary evolves to a giant, potentially filling then exceeding the volume of its Roche lobe. During this phase, dissipation associated with the very strong tides are expected to damp any orbital eccentricity and spin the giant's envelope close to synchronism with the orbit. The mass transfer that ensues can be stable or unstable. In either case, the interaction removes the envelope of the giant and leaves behind the core, which cools to become the WD. 

Whether the mass transfer from the red giant to the main sequence star is stable or unstable has dramatic consequences for the subsequent evolution. 
The stability of mass transfer depends on the binary mass ratio and the response of the giant star to mass loss. 
If the Roche lobe of the giant grows {\it relative} to the giant as mass is exchanged, then the mass transfer will be stable. 

\citet{1995MNRAS.273..731R} have described the outcome of these instances of stable mass transfer with a simple relation between the final orbital period and the WD mass. This is possible because the shell burning luminosity of a giant depends primarily on the compactness of the star's degenerate core. As a result, there is a well-defined relationship between core mass and giant envelope radius. A tight relation between eventual WD mass and binary orbital period emerges for systems formed through stable mass transfer interactions \citep{1995MNRAS.273..731R,2010ApJ...715...51V,2011ApJ...728..139C,2012ApJ...748..115B,2015ApJ...803...82R,2015ApJ...815...26F}.

If mass transfer is unstable, the mass transfer rate undergoes a runaway, eventually reaching rates much faster than the rate at which material can cool and accrete onto the main sequence star. In this case, the giant's envelope gas subsumes the binary pair, and forms a shared, common envelope around the two stars \citep{1976IAUS...73...75P}. Relative velocity between the cores and envelope gas gives rise to dynamical friction forces that drive the stars together on a dynamical timescale. If the deposition of orbital energy is sufficient to clear the surroundings of the binary, the remnant system of a WD and main sequence star, the post common envelope binary (PCEB), will exhibit an orbit transformed to much closer separation by this interaction \citep{1976IAUS...73...75P,1993PASP..105.1373I,2013AaARv..21...59I}.

\subsection{SLBs on the $P$--$M_{\rm WD}$ Plane}

Figure \ref{fig:pm} plots three SLBs reported in this paper (star symbols) with other known WD--MS binaries on $P$--$M_{\rm WD}$ plane. We also show the theoretical $P$--$M_{\rm WD}$ relation for the stable mass transfer case discussed in the previous subsection \citep{1995MNRAS.273..731R} and its updated version for a smaller mass \citep{2011ApJ...732...70L}. These lines set upper boundaries for post-interaction systems, below which the WD progenitor has likely experienced Roche-lobe overflow onto the MS companions.

The first discovered self-lensing binary, KOI-3278, has a significantly shorter period than this boundary. \citet{2014AaA...568L...9Z} studied the evolution of this system and found that it can be interpreted as a PCEB with progenitor masses of $M_{\mathrm{RG,i}}=2.45 M_\odot$ and $M_{\mathrm{MS,i}}=1.034 M_\odot$ and an initial orbital period of $1300\,\mathrm{days}$ (though we caution that the uncertainty in this estimate is much larger than the digits alone would suggest). Likewise, the relatively massive binary IK Peg and the WDs with red-dwarf companions mainly from SDSS are located well below the $P$--$M_{\rm WD}$ relation for the stable mass transfer scenario, and are also interpreted as PCEBs whose orbits were dramatically shrunk during the common envelope evolution \citep{2007MNRAS.382.1377R}.

Our SLBs demonstrate a further diversity of the post-interaction systems, populating the currently unexplored, long-period regime {(but see also Figure \ref{fig:fbs} and Section \ref{ssec:bs})}. Their wide orbits suggest that their orbits have not been tightened significantly via common-envelope evolution. {It appears possible that post-interaction binaries can lie anywhere between the densely-populated PCEB area and the stable mass transfer line}. 
{The WD masses of roughly $0.6\,M_\odot$ and orbital periods of SLBs seem consistent with stable, ``case C", mass transfer, in which the donor star is in the AGB phase and has a mainly convective envelope.} That said, the interpretation is not so clear-cut at the moment, considering that the WD mass uncertainty is essentially dominated by that of the primary mass from KIC, which is of limited reliability. In addition, the likely detection of non-zero eccentricities up to $\simeq0.2$, may not fit to the naive expectation from mass transfer either (Section \ref{ssec:eccentricity}). Understanding their detailed evolution history will be an important step toward understanding the possible outcomes of binary interactions in general. 

The self-lensing sample also seems to be a population distinct from other compact ($P<10\,\mathrm{days}$) eclipsing WDs with earlier-type companions from WASP (triangles) and {\it Kepler} (squares), although the lack of intermediate-period population may partly be due to the cancellation of the eclipse and self-lensing pulse --- as shown by the two dotted boundaries where the pulse height is twice larger than the eclipse depth (labeled as ``lensing dominant") and where the opposite is the case (``eclipse dominant"). The difference in the orbital separation by two orders of magnitude also indicates that the evolutionary phase of the WD progenitor during mass transfer was likely very different. Considering the {eclipse probability proportional to $P^{-2/3}$, the occurrence rate of long-period WD binaries as we found (roughly a few $0.1\%$ or more) seems much higher than that of} the short-period population. Such differences may reflect the relative abundance of natal binaries as a function of orbital period and spectral types, as well as the outcome of interaction. 

\subsection{Conditions for Stable Mass Transfer}\label{ssec:stablemt}

The current orbital separation and WD mass of SLB 1 (and perhaps SLBs 2 and 3 as well) are close to those expected from stable mass transfer, as is also the case for other compact systems with hot-dwarf companions. Here we discuss whether these systems satisfy minimal analytical requirements for this possible formation path.

For stable mass transfer to occur in a given binary with $q_\mathrm{i}=M_\mathrm{MS,i}/M_\mathrm{RG,i}$:
\begin{enumerate}
\item The progenitor of the WD must be more massive than that of the primary so that the former evolves faster; this condition translates into $q_{\rm i}<1$.
\item $q_{\rm i}$ must be larger than the critical value for the stable mass transfer, $q_{\rm crit}$, which is bound by Eqns. (\ref{eq:qcritconsv}) and (\ref{eq:qcritnonconsv}); see Appendix \ref{sec:mt_stability} for more details of this condition.
\item The progenitor of the WD must be sufficiently massive so that it can evolve into a giant within the cosmic age; this requires $M_{\rm RG,i} \ge m_{\rm H} M_\odot$ with $m_{\rm H}\approx0.9$ \citep{2016AaA...594A..13P}.
\end{enumerate}
Assuming that a fraction $\alpha(>0)$ of mass is lost during the transfer, the third condition is related to the observed total mass $M_{\rm tot}=(1-\alpha)(M_{\rm MS,i}+M_{\rm RG,i})$ via
\begin{equation}
	\label{eq:qinit_mtot}
	\frac{M_{\rm tot}}{1-\alpha}=(1+q_{\rm i}) M_{\rm RG, i}\geq(1 + q_{\rm i}) m_{\rm H} M_\odot.
\end{equation}

\begin{figure*}[tbp]
\begin{center}
\includegraphics[bb=0 0 919 625,width=0.7\linewidth]{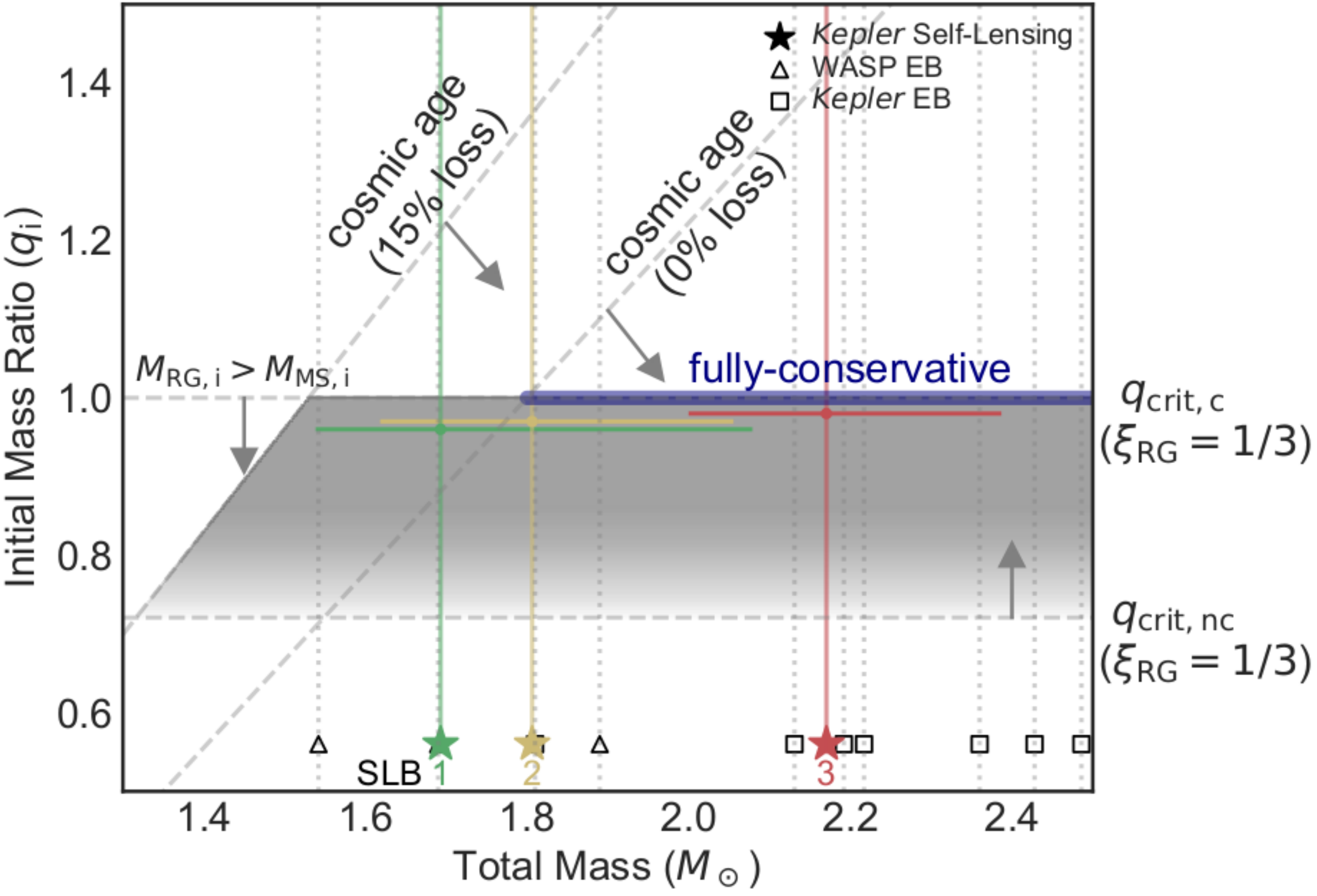}
\caption{The constraint on the initial mass ratio and total mass in the stable mass transfer scenario. Each symbol at the bottom corresponds to the WD binary system whose position on the $P$--$M_{\rm WD}$ plane is consistent with this scenario: SLBs (stars), {\it Kepler} EBs (squares), and WASP EBs (triangles). The horizontal bars show the uncertainty of the total mass of the SLBs. The region consistent with partially non-conservative, stable mass transfer (with the total mass lost by $15\%$ in this example) is shaded with gray. The thick navy blue line indicates the same solution for the fully-conservative case (i.e. $0\%$ mass loss). See the text for details.
 \label{fig:mtot}}
\end{center}
\end{figure*}

These conditions for stable mass transfer are illustrated as the gray-shaded region in Figure \ref{fig:mtot}, for $\alpha=0.15$ and $q_{\rm crit, nc}\approx0.72$ corresponding to $\xi_{\rm RG}=1/3$ (Eqn. \ref{HW}). While the actual value of $q_{\rm crit, nc}$ for a partly non-conservative case is likely larger than Eqn.~\ref{eq:qcritnonconsv}, the difference does not affect the discussion here as long as $q_{\rm crit, nc}<1$. 
The vertical lines denote the observed total masses of the WD binary systems (including SLBs 1--3 shown with horizontal $1\sigma$ error bars) close to the stable mass transfer prediction in Figure \ref{fig:pm}. The overlap between the lines and the shaded region shows that all of these systems satisfy the necessary conditions for stable mass transfer if $\alpha\gtrsim 0.15$. 

We also draw the third condition for the fully-conservative transfer ($\alpha=0$; ``0\% loss" line), along with $q_{\rm crit,c}=1$ for $\xi_{\rm RG}=1/3$. In this case, the only allowed region would be $q_{\rm i}\simeq1$ and total mass $\gtrsim 1.8\,M_\odot$, as indicated by the thick navy blue line. Again most of the systems qualify, at least in terms of the estimated total mass of the system.

\subsection{Implications of Possible Orbital Eccentricity} \label{ssec:eccentricity}

Binary systems are formed with a wide range of eccentricities. On the main sequence, pairs of roughly solar mass stars have a broad eccentricity distribution \citep[][Figure 32]{2017ApJS..230...15M}. As one of the component stars fills its Roche lobe it is strongly distorted by the tidal force from its companion. The imprint of this tidal forcing on turbulent motion in convective giant-star envelopes provides a means where net work is translated from the orbit to the stellar envelope material \citep[e.g.][]{2008EAS....29...67Z,2014ARA&A..52..171O}. This tidal dissipation acts to spin up the envelope into synchronous rotation with the orbital motion and to damp orbital eccentricity. 

Tides, and in particular their associated dissipation rates, act as a strong function of distance between two bodies. The synchronization time scales as $(R_\star/a)^6$, while the circularization time scales as $(R_\star/a)^8$ \citep{2008EAS....29...67Z}.  Because of this strong scaling, it is generally thought that when systems overflow their Roche lobes and transfer mass to a companion, relative spin and orbital eccentricity should be damped to near-zero values. 

It is also worth noting that the circularization time is typically much longer than the synchronization time, because for most binary mass ratios, the inertia of the orbit is much greater than the inertia of the giant star's envelope \citep{2008EAS....29...67Z}. Thus a binary system may easily be (pseudo) synchronous while maintaining non-zero orbital eccentricity.  As a result, an exception to the expectation of low eccentricity may arise when there is insufficient time during the interaction phase for tidal dissipation to reduce the eccentricity from its original value.  Such an argument is most frequently invoked for systems that do not fully fill their Roche lobe, such as the wind-fed Symbiotic stars \citep[e.g.][]{1986syst.book.....K}.
If the Roche filling phase in the formation of SLBs were short-lived (less than a few times the circularization timescale) this could offer a possible explanation of the low, but non-zero, detected eccentricities.

Another possible reason for eccentricity in the present SLBs is an excitation of eccentricity post interaction.  In clusters, where binary-single star encounters are frequent (particularly compared to the $\sim$Gyr lifetimes of the SLBs), mild pumping of orbital eccentricities is thought to be common \citep{2011Natur.478..356G,2015ebss.book..251P}. Another possible channel of eccentricity excitation is secular interaction with a tertiary companion to an (otherwise isolated) binary. Under this scenario, we would expect post interaction eccentricity in a fraction of systems similar to the fraction which are, in fact, higher order multiples. For solar-mass stars, this fraction is $\sim$10\%, so detection of eccentricity in multiple of the SLBs is somewhat suprising. The presence of (or constraints on the properties of) a tertiary component may be testable with long-term RV monitoring of SLBs. 

By filling in a new parameter space in the $P$--$M_{\rm WD}$ diagram, the detailed properties of the orbits of SLBs will offer a window into the mass transfer and dissipation processes that lead to evolution of not just the components of the binary system, but also their orbit.

\subsection{{Connections between SLBs and Blue Stragglers}}\label{ssec:bs}

\begin{figure*}[htbp]
\begin{center}
\includegraphics[width=\linewidth,bb=0 0 720 432]{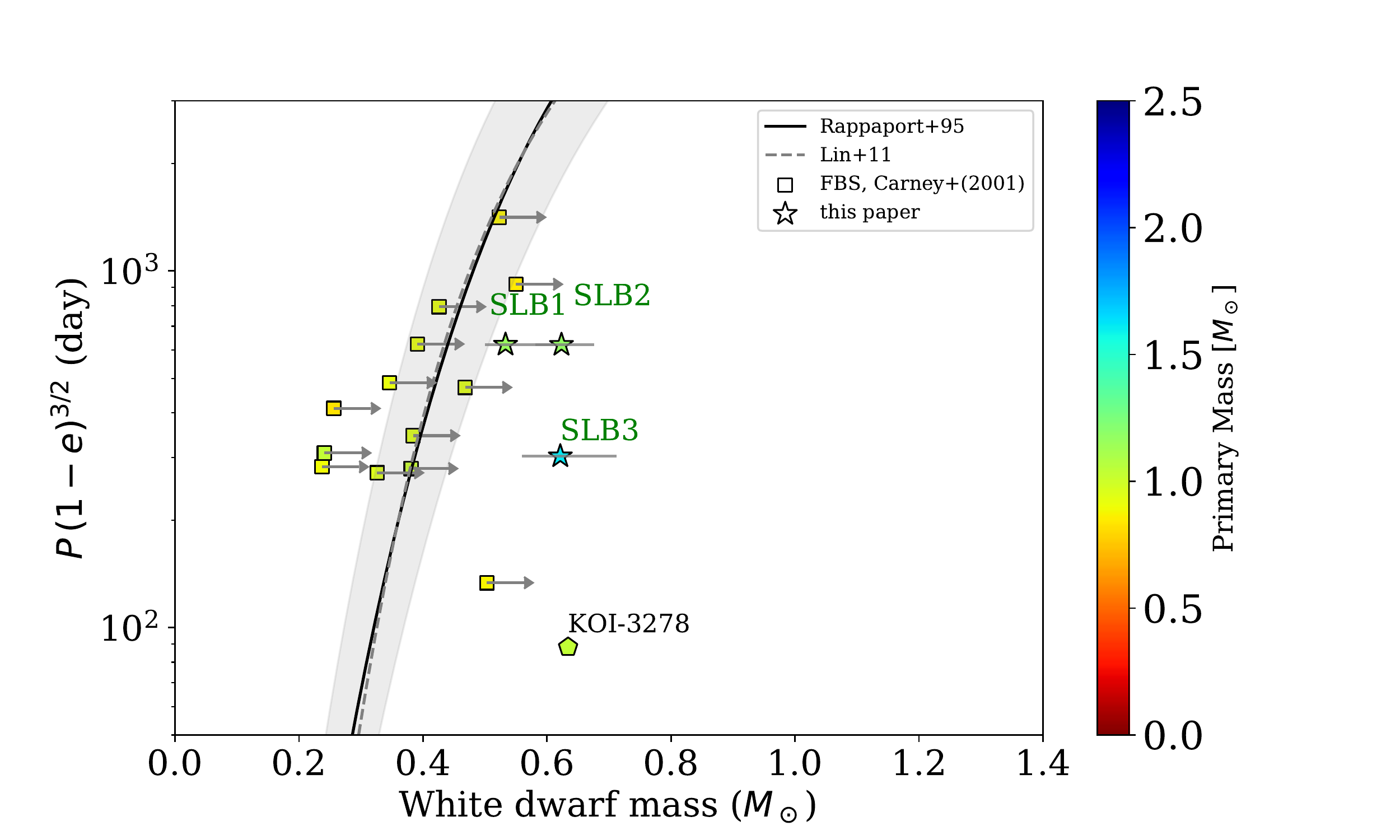}
\caption{{Field blue straggler binaries and SLBs in the $P$--$M_{\rm WD}$ plane. The orbital periods and minimum masses of field blue straggler binaries (square symbols) are adopted from Tables 4 and 5 in \citet{2001AJ....122.3419C}. The star symbols represent the SLBs reported in this paper and the pentagon symbol denotes KOI-3278. The solid and dashed lines and the shaded region are the same as those in Figure \ref{fig:pm}.  The $y$-axis indicates the period corresponding to the separation at periastron, $P(1-e)^{3/2}$, instead of $P$. \label{fig:fbs}}}
\end{center}
\end{figure*}

{
Blue stragglers are cluster stars so named because they extend beyond the normal (single age) isochrone in a star cluster's color--magnitude diagram \citep{1953AJ.....58...61S,1955ApJ...121..616J}. 
Mass transfer in a binary has been proposed to be one of their formation paths, with the other possibilities being mergers and collisions.
For example, the mass function of spectroscopic companions to blue stragglers in the open cluster NGC 188 places most of the companions around $0.5M_\odot$ with orbital periods grouped within a factor of a few $1000$~days \citep{2009Natur.462.1032M,2011Natur.478..356G}, suggesting that they are WD remnants of stable mass transfer. Indeed, some exhibit UV excess \citep{2014ApJ...783L...8G} suggestive of emission from a hot WD surface. Blue straggler binaries in the open cluster M67 also show similar orbital properties \citep{2007HiA....14..444L}.
}

{Such unusually blue stars have also been identified in old thick disk and halo populations of the Galaxy, defined by high proper motions and low metallicities \citep{2000AJ....120.1014P, 2001AJ....122.3419C}. These hot and massive field stars, as implied by their blue colors, appear to be too young to belong to their parent old populations. They are thus interpreted as analogs of the blue stragglers in clusters. These ``field blue stragglers" (FBSs) are frequently in single-lined spectroscopic binaries with similar characteristics to a significant subset of blue straggler binaries in clusters, for which mass-transfer origin has been suggested. FBSs are better suited for studying blue stragglers formation via mass transfer, since the contributions from other formation paths are expected to be small in the field.} 

{SLBs 1--3 are likely field binaries with very similar characteristics to the FBS binaries as discussed above (Figure \ref{fig:fbs}), and appear to be an eclipsing subset of this classical FBS population.
They provide further evidence for the mass transfer origin of FBSs by directly showing that the companions in such binaries are actually WDs as remnants of mass transfer.   
Moreover, the self lensing provides a better means to identify and study the products of mass transfer in general. The method is not limited to stars with extreme kinematics and metallicities, unlike the FBS binaries studied so far, and yields the actual WD masses without ambiguity of the orbital inclination. }

\section{Summary and Conclusion}\label{sec:summary}

We have discovered three self-lensing binaries in the {\it Kepler} data. The pulse light curves and RVs are consistent with binaries of low-mass stars with WD companions of $M_{\rm WD}\simeq0.6\,M_\odot$ in wide ($P=1$--$2\,\mathrm{yr}$) and low-eccentricity ($e\lesssim0.2$), edge-on orbits. The absence of secondary eclipses implies relatively cool WDs with $T\lesssim10^{4}\,\mathrm{K}$. The inferred WD masses and orbit separations imply that the WD progenitors have transferred their masses onto the stellar primaries in the past, but without leading to common-envelope evolution as suggested for other shorter-period WD--MS binaries, including the first SLB system KOI-3278.

Our self-lensing sample populates the longest-period regime of the $P$--$M_{\rm WD}$ plane. The sample allows us to probe the diversity of the post-interaction systems, ranging from short-period PCEBs originating from unstable mass transfer to longer-period systems that could be the outcome of stable mass transfer. If more precisely characterized with follow-up observations, they may provide stringent constraints on the conditions that lead to the occurrence of common envelope phases in binary systems. 
In addition, the SLBs reported in this paper carry many similarities to blue stragglers with WD companions for which mass-transfer origin has been suggested. Thus, these SLBs may also be ideal environments for better understanding the origin of blue stragglers as their field analogs.

\acknowledgments

This publication makes use of VOSA developed under the Spanish Virtual Observatory project supported from the Spanish MICINN through grant AyA2011-24052. We thank Josh Winn, Saul Rappaport, Jennifer Sokoloski, Scott Kenyon, and Toshikazu Shigeyama for helpful conversations. H.K. is supported by Grant-in-Aid for Young Scientists (B) from Japan Society for Promotion of Science, No.\,17K14246 and the Astrobiology Center from NINS. This work was performed in part under contract with the California Institute of Technology/Jet Propulsion Laboratory funded by NASA through the Sagan Fellowship Program executed by the NASA Exoplanet Science Institute. This work was partly supported by the JSPS Core-to-Core Program ``Planet$^2$" and the Einstein Fellowship Program executed by the Smithsonian Astrophysical Observatory.

\appendix

\section{Additional Vetting of the Self-lensing Candidates}

\subsection{Centroid Analysis of Pixel-Level Difference Imaging} \label{ssec:centroid}

The centroid offset of the difference image has been used as a good indicator of the contamination of the neighboring stars, especially for weak transit signals \citep{2013PASP..125..889B}. The difference image is usually computed by subtracting the mean in-transit image from the out-of-transit one. Because the pulse increases the flux, we compute the difference image with the opposite sign to that of a planetary transit:
\begin{eqnarray}
  \delta f(x,y) \equiv \langle f(x,y,t) \rangle_{\mathrm{in}} - \langle f(x,y,t) \rangle_{\mathrm{out}},
\end{eqnarray}
where $x$ and $y$ are the pixel coordinates, $f(x,y,t)$ is the flux at the pixel $(x,y)$ as a function of time, and $\langle \rangle_{\mathrm{in}}$ and $\langle \rangle_{\mathrm{out}}$ are time-averages of $f(x,y,t)$ inside and outside of the pulse, respectively. The in-pulse flux was averaged over the pulse duration, while the out-of-pulse averaging was performed for the data on both sides of the pulse with a similar length to the pulse duration. The resulting difference image $\delta f(x,y)$ was fitted with the point spread function model using the {\it kepprf} routine in the PyKE package \citep{2012ascl.soft08004S} to derive its centroid.  
In the same way we also compute the centroid for the mean image of the quarter after masking the pulses, and give the centroid offset as the difference between the two.  

\citet{2013PASP..125..889B} used the mean of the centroid offsets for all the transits and its variance to estimate the statistical significance. Because we have only two or three pulses, we estimate the statistical significance of the offset by analyzing simulated pulse signals. We artificially inject pulses to the randomly sampled parts of the out-of-pulse light curves (67 for each season) and analyze their centroid offsets in the same manner as above. The scatter of these simulated offsets is interpreted as the statistical distribution of the uncertainty. 

\begin{figure*}[htbp]
\begin{center}
\includegraphics[bb=0 0 1216 789,width=0.95\linewidth]{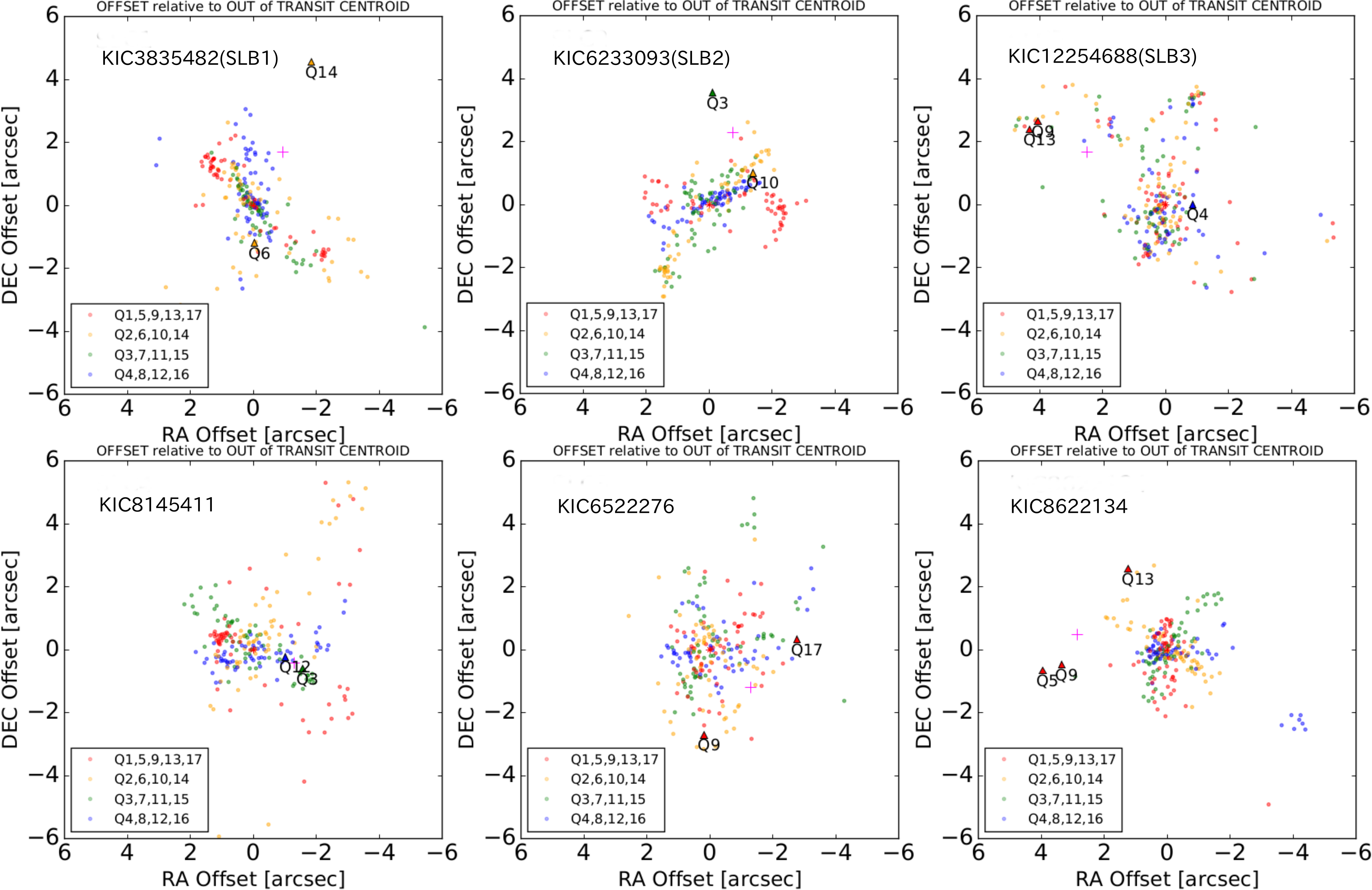}
\caption{Centroid offsets of the difference images for the candidates listed in Table \ref{tab:candidates}. The triangles show the offsets for the observed pulses, whose means are shown by purple crosses. The colored circles are the offsets computed for the simulated mock pulses, whose scatter is interpreted as the statistical uncertainty. The four colors correspond to four different seasons. 
  \label{fig:pixel}}
\end{center}
\end{figure*}

Figure \ref{fig:pixel} displays the centroid offsets of the observed pulses and those of the mock pulses (to represent the statistical uncertainty). The scatter in the mock offsets strongly depends on the season and is far from Gaussian for some quarters. This tendency is stronger than in the offsets of typical KOIs, and is likely attributed to the local trend in the light curve: owing to their long orbital periods, our candidates have longer pulse durations than typical KOIs, and therefore the difference images are more sensitive to the trend. 

The offsets of the pulses are consistent with the statistical uncertainty except for KIC 3835482 (Q14), KIC 6233093 (Q3), and the three pulses of KIC 8622134. The first two outliers are not the obvious signature of contamination, because they can be explained by the gaps in the left side of the pulses (see Figure \ref{fig:allp}); when either of the two sides of the signal is missing, the offset becomes sensitive to the local linear trend \citep{2013PASP..125..889B}. On the other hand, the centroid shift of KIC 8622134 is problematic. While the observed shifts are not so large compared to the overall scatter of the mock offsets, the shifts during the pulses show significant deviation if we focus on the points in the same season as the pulses (i.e., red points for Q1, 5, 9, 13, and 17). This suggests that the pulses of KIC 8622134 are likely due to contamination, as also indicated by the ephemeris matching test (Section \ref{ssec:ephemeris}).

\subsection{Comparison with Other Known Pulses}

\begin{figure*}[htbp]
\begin{center}
\includegraphics[bb=0 0 1080 216,width=0.9\linewidth]{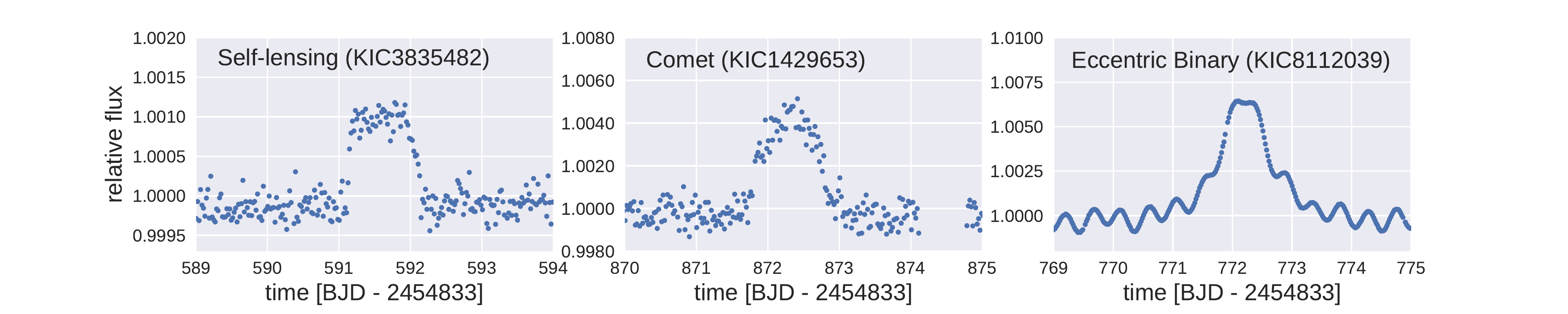}
\caption{Comparison of typical symmetric pulses in the {\it Kepler} light curves (PDCSAP). Left: Self-lensing binary (SLB1) we found. Middle: Comet C/2007 Q3 \citep{2014ApJ...786..158G}. Right: eccentric binary \citep{2011ApJS..197....4W}. \label{fig:exa}}
\end{center}
\end{figure*}

In Figure \ref{fig:exa}, we compare the signals we found with other known sources of pulses. Reflected light from asteroids/comets induces a single pulse in the {\it Kepler} light curves \citep{2014ApJ...786..158G}. The middle panel in Figure \ref{fig:exa} displays a symmetric pulse by a comet, though most of them have an asymmetric shape. The comet pulse can be a false positive for a single self-lensing event with a low S/N. However, this cannot be the source of such repeating and symmetric pulses as discussed in this paper.

An eccentric binary with a very small pericenter distance can also produce periodic and symmetric pulses by tidal deformation of a star. The right panel in Figure \ref{fig:exa} shows the famous example of such a ``heartbeat" eccentric binary, KIC 8112039 \citep{2011ApJS..197....4W}. Although the pulse is V-shaped rather than top-hat as expected for the self-lensing, the signals produced by such heartbeat binaries are strictly periodic and so can be a false positive for the low S/N cases. As shown in Figure \ref{fig:allp}, the pulses of KIC 3835482, KIC 6233093, and KIC 12254688 clearly exhibit top-hat shapes, and thus are unlikely to be the heartbeat binaries. However, the heartbeat scenario may not be fully excluded for KIC 6522276 and KIC 8145411 with low S/N {based on the light curve alone. Thus we classified KIC 8145411 to be an unconfirmed candidate, since the orbit is not yet well constrained from RVs. KIC 6522276, on the other hand, did not show any significant RV variations (Section \ref{ssec:vetting_rv}), which is inconsistent with either of the self-lensing or eccentric binary scenario. The pulse signal was thus classified to be a false positive.}

\section{Search for Asteroseismic Oscillations} \label{ssec:seismology}

Although the primaries of our SLBs have $Kp\gtrsim13$, they might still be sufficiently bright to detect solar-like pulsations from a star in the red giant phase \citep{Mathur2016}. Thus stellar pulsations were searched in the power spectra of the PDCSAP long cadence data. Before computing the spectra, we corrected for the jumps between the successive quarters and removed the spurious data points in a similar fashion as in \citet{Hirano2015} or \citet{Garcia2011}. 

A visual inspection did not allow us to detect any signal compatible with solar-like oscillations within the available frequency range ($0$ to $\approx 270\,\mu$Hz). Although this analysis does not exclude that the SLBs pulsate at a higher frequency typical for sub-giants or main-sequence stars, the null detection at low frequency suggests that those stars probably have not reached the red-giant phase.

\section{Mass--radius relations of white dwarfs} \label{sec:mrrelation}

\begin{figure}[htbp]
\begin{center}
\includegraphics[bb=0 0 389 266,width=0.6\linewidth]{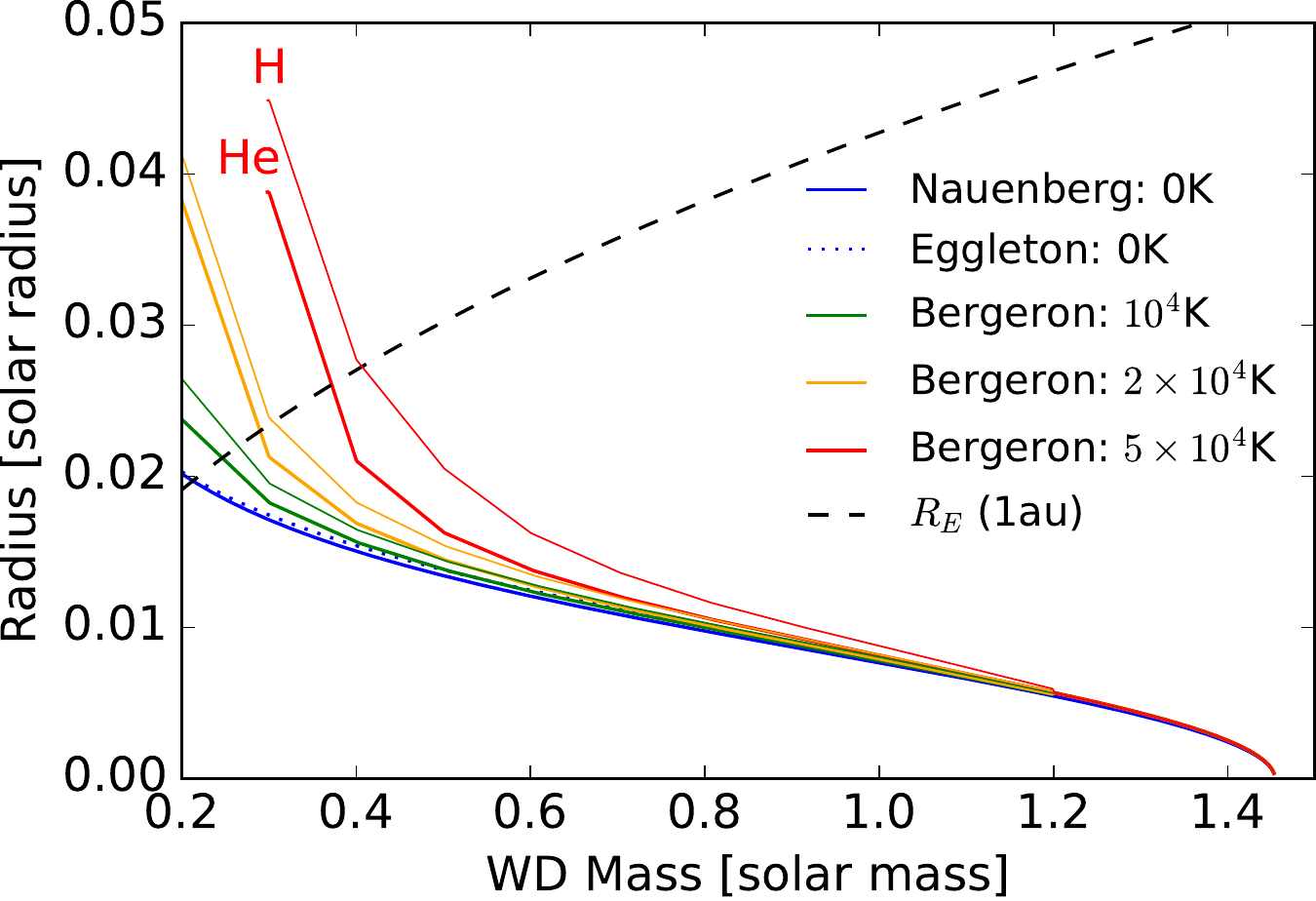}
\caption{The mass--radius relations of the white dwarf. Each color indicates different temperature of the white dwarf: $0$~K (blue),  $10,000$~K (green), $20,000$~K (yellow), and $50,000$~K (red). The thin and thick lines correspond to the pure-hydrogen and pure-helium atmospheres, respectively. 
The black dashed line is the Einstein radius for the semi-major axis of 1~au. \label{fig:MR}}
\end{center}
\end{figure}

Figure \ref{fig:MR} displays the mass--radius relations of white dwarfs. The Nauenberg relation \citep{1972ApJ...175..417N} is the mass--radius relation for the zero-temperature white dwarf:
\begin{eqnarray}
  \frac{R_\mathrm{WD}}{R_\odot} = 0.0225 \mu^{-1} \left[ \left( \frac{\Mwd}{\Mch} \right)^{-2/3} - \left( \frac{\Mwd}{\Mch} \right)^{2/3} \right]^{1/2},
\end{eqnarray}
where $\Mwd$ is the white dwarf mass, $\Mch = 1.454 M_\odot$ is the Chandrasekhar mass, and $\mu$ is the number of nucleons per electron (we adopt $\mu=2$). We also plot an alternative mass--radius relation for the zero-temperature, referred to as the Eggleton relation \citep{1988ApJ...332..193V}:
\begin{eqnarray}
  \frac{R_\mathrm{WD}}{R_\odot} = 0.0225 \mu^{-1} \left[ \left( \frac{\Mwd}{\Mch} \right)^{-2/3} - \left( \frac{\Mwd}{\Mch} \right)^{2/3} \right]^{1/2} \left[ 1+ 3.5 \left( \frac{\Mwd}{\Mch} \right)^{-2/3} + \left( \frac{\Mwd}{\Mch} \right)^{-1} \right]^{-2/3},
\end{eqnarray}
although the Eggleton relation is not significantly different from the Nauenberg relation. The dashed line is the Einstein radius for $a=1$~au given by Eqn.~(\ref{eq:re}). This indicates that the dimming due to an eclipse becomes important for WDs $\lesssim0.4M_\odot$, as the Einstein radius shrinks and the WD radius increases.

In Figure \ref{fig:MR}, we also show the mass--radius relations for different effective temperatures from the WD cooling model for the pure-helium (DB; thick lines) and the pure-hydrogen (DA; thin lines) atmospheres provided by Pierre Bergeron\footnote{http://www.astro.umontreal.ca/~bergeron/CoolingModels} \citep{2006AJ....132.1221H,2006ApJ...651L.137K,2011ApJ...730..128T,2011ApJ...737...28B}. 
For $T \lesssim 3500$~K (pure-helium) or $T \lesssim 2500$~K (pure-hydrogen), the relations converge to the Nauenberg and Eggleton relations. 
These two models were used in Section \ref{ssec:lc_luminosity} to derive constraints on the WD age.

\section{Estimating the Stability of Mass Transfer}\label{sec:mt_stability}

The stability of mass transfer in a binary system depends on the relative response of the donor star and its Roche lobe. 
We therefore  compare the Roche lobe response to mass transfer,
$\xi_{\mathrm{RL}} \equiv \left( d \ln{R_{\mathrm{RL}}} / d \ln{M_{\mathrm{RG}}} \right)$,
to the response of the giant,
$\xi_{\mathrm{RG}} \equiv \left( d \ln{R_{\mathrm{RG}}} / d \ln{M_{\mathrm{RG}}} \right)$,
where $R_{\rm RL}$ is the radius of the Roche lobe, $R_{\rm RG}$ is the radius of the giant and $M_{\rm RG}$ is the mass of the giant. If $\xi_{\mathrm{RL}}< \xi_{\mathrm{RG}}$, then mass transfer remains stable. If $\xi_{\mathrm{RL}}> \xi_{\mathrm{RG}}$, then mass transfer unstably increases.

The Roche lobe response, $\xi_{\mathrm{RL}}$, depends on the binary mass ratio and the stellar response, $\xi_{\mathrm{RG}}$, depends on the giant star's structure and the timescale of mass transfer. For rapid mass exchange we can approximate $\xi_{\mathrm{RG}}$ with the adiabatic response of the star to changing mass,
$\xi_{\mathrm{ad}} \equiv \left( d \ln{R_{\mathrm{RG}}} /d \ln{M_{\mathrm{RG}}} \right)_s,$
which is strictly valid only in cases where the mass transfer rate is sufficiently high that radiative losses from the outer layers of the star are unimportant. 
Under these simplifications, the \citet{1987ApJ...318..794H} composite polytrope model predicts the following response for polytropic structures with a condensed core
\begin{eqnarray}\label{HW}
\xi_{\rm ad} \approx {1 \over 3-n} \left(1-n + {m_{c} \over 1-m_{c}} \right),
\end{eqnarray}
which, for example, gives $\xi_{\rm ad}=1/3$ for $n=1.5$ and $m_{c}=0.5$, parameters approximately relevant to a $1M_\odot$ red giant with a $0.5M_\odot$ core. However, the full applicability of this approximation is debated. For example, \citet{2015MNRAS.449.4415P} simulate mass loss from 1D stellar models and argue that in their simulation, low mass binaries might be mildly more stable than one would expect from the adiabatic approximation (see their figures 11 and 14, note that the effects are much larger for high mass donor stars). 

Given $\xi_{\rm RG}$, we can estimate the critical mass ratio that divides stable systems from unstable systems. Two limiting cases are when the mass transfer is fully conservative (the total system mass is constant) or fully non-conservative (mass transferred is lost from the system). 
For the mass conserved case,  stable mass transfer is possible for $q > q_\mathrm{crit,c}$ \citep{1995MNRAS.273..731R} with 
\begin{eqnarray}
\label{eq:qcritconsv}
  q_\mathrm{crit,c} = \left( {5 \over 6} + {\xi_{\mathrm{RG}} \over 2}   \right)^{-1}.
\end{eqnarray}
If the mass transfer is non-conservative, then stable mass transfer is possible for $q>q_\mathrm{crit,nc}$ \citep{1995MNRAS.273..731R}, where $q_\mathrm{crit,nc}$ is given by
\begin{eqnarray}
\label{eq:qcritnonconsv}
{q_\mathrm{crit,nc} + 3 \over 3(1+q_\mathrm{crit,nc}) q_\mathrm{crit,nc}} = {5 \over 6} + {\xi_{\mathrm{RG}} \over 2}.
\end{eqnarray}
The value of $q_\mathrm{crit,nc}$ is generally smaller than  $q_\mathrm{crit,c}$. Therefore, a wider range of binaries can achieve stable transfer in the non-conservative limit.

If the initial binary mass ratio, $q_\mathrm{i}= M_\mathrm{MS,i}/M_\mathrm{RG,i}>q_\mathrm{crit}$, the mass transfer becomes stable.

\bibliographystyle{aasjournal}

\begin{thebibliography}{}
\expandafter\ifx\csname natexlab\endcsname\relax\def\natexlab#1{#1}\fi
\providecommand{\url}[1]{\href{#1}{#1}}

\bibitem[{{Agol}(2002)}]{2002ApJ...579..430A}
{Agol}, E. 2002, Astrophysical Journal, 579, 430

\bibitem[{{Agol}(2003)}]{2003ApJ...594..449A}
---. 2003, Astrophysical Journal, 594, 449

\bibitem[{{Bayo} {et~al.}(2008){Bayo}, {Rodrigo}, {Barrado Y Navascu{\'e}s},
  {Solano}, {Guti{\'e}rrez}, {Morales-Calder{\'o}n}, \&
  {Allard}}]{2008AaA...492..277B}
{Bayo}, A., {Rodrigo}, C., {Barrado Y Navascu{\'e}s}, D., {et~al.} 2008,
  Astronomy and Astrophysics, 492, 277

\bibitem[{{Bergeron} {et~al.}(2011){Bergeron}, {Wesemael}, {Dufour},
  {Beauchamp}, {Hunter}, {Saffer}, {Gianninas}, {Ruiz}, {Limoges}, {Dufour},
  {Fontaine}, \& {Liebert}}]{2011ApJ...737...28B}
{Bergeron}, P., {Wesemael}, F., {Dufour}, P., {et~al.} 2011, Astrophysical
  Journal, 737, 28

\bibitem[{{Beskin} \& {Tuntsov}(2002)}]{2002AaA...394..489B}
{Beskin}, G.~M., \& {Tuntsov}, A.~V. 2002, Astronomy and Astrophysics, 394, 489

\bibitem[{{Bloemen} {et~al.}(2011){Bloemen}, {Marsh}, {{\O}stensen},
  {Charpinet}, {Fontaine}, {Degroote}, {Heber}, {Kawaler}, {Aerts}, {Green},
  {Telting}, {Brassard}, {G{\"a}nsicke}, {Handler}, {Kurtz}, {Silvotti}, {Van
  Grootel}, {Lindberg}, {Pursimo}, {Wilson}, {Gilliland}, {Kjeldsen},
  {Christensen-Dalsgaard}, {Borucki}, {Koch}, {Jenkins}, \&
  {Klaus}}]{2011MNRAS.410.1787B}
{Bloemen}, S., {Marsh}, T.~R., {{\O}stensen}, R.~H., {et~al.} 2011, Monthly
  Notices of the Royal Astronomical Society, 410, 1787

\bibitem[{{Breton} {et~al.}(2012){Breton}, {Rappaport}, {van Kerkwijk}, \&
  {Carter}}]{2012ApJ...748..115B}
{Breton}, R.~P., {Rappaport}, S.~A., {van Kerkwijk}, M.~H., \& {Carter}, J.~A.
  2012, Astrophysical Journal, 748, 115

\bibitem[{{Bryson} {et~al.}(2013){Bryson}, {Jenkins}, {Gilliland}, {Twicken},
  {Clarke}, {Rowe}, {Caldwell}, {Batalha}, {Mullally}, {Haas}, \&
  {Tenenbaum}}]{2013PASP..125..889B}
{Bryson}, S.~T., {Jenkins}, J.~M., {Gilliland}, R.~L., {et~al.} 2013,
  Publications of the Astronomical Society of the Pacific, 125, 889

\bibitem[{{Carney} {et~al.}(2001){Carney}, {Latham}, {Laird}, {Grant}, \&
  {Morse}}]{2001AJ....122.3419C}
{Carney}, B.~W., {Latham}, D.~W., {Laird}, J.~B., {Grant}, C.~E., \& {Morse},
  J.~A. 2001, Astronomical Journal, 122, 3419

\bibitem[{{Carter} {et~al.}(2011){Carter}, {Rappaport}, \&
  {Fabrycky}}]{2011ApJ...728..139C}
{Carter}, J.~A., {Rappaport}, S., \& {Fabrycky}, D. 2011, Astrophysical
  Journal, 728, 139

\bibitem[{{Faigler} {et~al.}(2015){Faigler}, {Kull}, {Mazeh}, {Kiefer},
  {Latham}, \& {Bloemen}}]{2015ApJ...815...26F}
{Faigler}, S., {Kull}, I., {Mazeh}, T., {et~al.} 2015, Astrophysical Journal,
  815, 26

\bibitem[{{Farihi} {et~al.}(2010){Farihi}, {Hoard}, \&
  {Wachter}}]{2010ApJS..190..275F}
{Farihi}, J., {Hoard}, D.~W., \& {Wachter}, S. 2010, Astrophysical Journals,
  190, 275

\bibitem[{{Farmer} \& {Agol}(2003)}]{2003ApJ...592.1151F}
{Farmer}, A.~J., \& {Agol}, E. 2003, Astrophysical Journal, 592, 1151

\bibitem[{{Foreman-Mackey} {et~al.}(2013){Foreman-Mackey}, {Hogg}, {Lang}, \&
  {Goodman}}]{2013PASP..125..306F}
{Foreman-Mackey}, D., {Hogg}, D.~W., {Lang}, D., \& {Goodman}, J. 2013,
  Publications of the Astronomical Society of the Pacific, 125, 306

\bibitem[{{Foreman-Mackey} {et~al.}(2016){Foreman-Mackey}, {Morton}, {Hogg},
  {Agol}, \& {Sch{\"o}lkopf}}]{2016AJ....152..206F}
{Foreman-Mackey}, D., {Morton}, T.~D., {Hogg}, D.~W., {Agol}, E., \&
  {Sch{\"o}lkopf}, B. 2016, Astronomical Journal, 152, 206

\bibitem[{{Garc{\'{\i}}a} {et~al.}(2011){Garc{\'{\i}}a}, {Hekker}, {Stello},
  {Guti{\'e}rrez-Soto}, {Handberg}, {Huber}, {Karoff}, {Uytterhoeven},
  {Appourchaux}, {Chaplin}, {Elsworth}, {Mathur}, {Ballot},
  {Christensen-Dalsgaard}, {Gilliland}, {Houdek}, {Jenkins}, {Kjeldsen},
  {McCauliff}, {Metcalfe}, {Middour}, {Molenda-Zakowicz}, {Monteiro}, {Smith},
  \& {Thompson}}]{Garcia2011}
{Garc{\'{\i}}a}, R.~A., {Hekker}, S., {Stello}, D., {et~al.} 2011, Monthly
  Notices of the Royal Astronomical Society, 414, L6

\bibitem[{{Geller} \& {Mathieu}(2011)}]{2011Natur.478..356G}
{Geller}, A.~M., \& {Mathieu}, R.~D. 2011, Nature, 478, 356

\bibitem[{{Gosnell} {et~al.}(2014){Gosnell}, {Mathieu}, {Geller}, {Sills},
  {Leigh}, \& {Knigge}}]{2014ApJ...783L...8G}
{Gosnell}, N.~M., {Mathieu}, R.~D., {Geller}, A.~M., {et~al.} 2014,
  Astrophysical Journal Letters, 783, L8

\bibitem[{{Gould}(1995)}]{1995ApJ...441...77G}
{Gould}, A. 1995, Astrophysical Journal, 441, 77

\bibitem[{{Griest} {et~al.}(2014){Griest}, {Cieplak}, \&
  {Lehner}}]{2014ApJ...786..158G}
{Griest}, K., {Cieplak}, A.~M., \& {Lehner}, M.~J. 2014, Astrophysical Journal,
  786, 158

\bibitem[{{Han}(2016)}]{2016ApJ...820...53H}
{Han}, C. 2016, Astrophysical Journal, 820, 53

\bibitem[{{Hauschildt} {et~al.}(1999){Hauschildt}, {Allard}, \&
  {Baron}}]{1999ApJ...512..377H}
{Hauschildt}, P.~H., {Allard}, F., \& {Baron}, E. 1999, Astrophysical Journal,
  512, 377

\bibitem[{{Hirano} {et~al.}(2015){Hirano}, {Masuda}, {Sato}, {Benomar},
  {Takeda}, {Omiya}, {Harakawa}, \& {Kobayashi}}]{Hirano2015}
{Hirano}, T., {Masuda}, K., {Sato}, B., {et~al.} 2015, Astrophysical Journal,
  799, 9

\bibitem[{{Hjellming} \& {Webbink}(1987)}]{1987ApJ...318..794H}
{Hjellming}, M.~S., \& {Webbink}, R.~F. 1987, Astrophysical Journal, 318, 794

\bibitem[{{Holberg} \& {Bergeron}(2006)}]{2006AJ....132.1221H}
{Holberg}, J.~B., \& {Bergeron}, P. 2006, Astronomical Journal, 132, 1221

\bibitem[{{Iben} \& {Livio}(1993)}]{1993PASP..105.1373I}
{Iben}, Jr., I., \& {Livio}, M. 1993, Publications of the Astronomical Society
  of the Pacific, 105, 1373

\bibitem[{{Ivanova} {et~al.}(2013){Ivanova}, {Justham}, {Chen}, {De Marco},
  {Fryer}, {Gaburov}, {Ge}, {Glebbeek}, {Han}, {Li}, {Lu}, {Marsh},
  {Podsiadlowski}, {Potter}, {Soker}, {Taam}, {Tauris}, {van den Heuvel}, \&
  {Webbink}}]{2013AaARv..21...59I}
{Ivanova}, N., {Justham}, S., {Chen}, X., {et~al.} 2013, Astronomy and
  Astrophysicsr, 21, 59

\bibitem[{{Johnson} \& {Sandage}(1955)}]{1955ApJ...121..616J}
{Johnson}, H.~L., \& {Sandage}, A.~R. 1955, Astrophysical Journal, 121, 616

\bibitem[{{Kenyon}(1986)}]{1986syst.book.....K}
{Kenyon}, S.~J. 1986, {The symbiotic stars}

\bibitem[{{Kipping}(2013)}]{2013MNRAS.435.2152K}
{Kipping}, D.~M. 2013, \mnras, 435, 2152

\bibitem[{{Kirk} {et~al.}(2016){Kirk}, {Conroy}, {Pr{\v s}a}, {Abdul-Masih},
  {Kochoska}, {Matijevi{\v c}}, {Hambleton}, {Barclay}, {Bloemen}, {Boyajian},
  {Doyle}, {Fulton}, {Hoekstra}, {Jek}, {Kane}, {Kostov}, {Latham}, {Mazeh},
  {Orosz}, {Pepper}, {Quarles}, {Ragozzine}, {Shporer}, {Southworth},
  {Stassun}, {Thompson}, {Welsh}, {Agol}, {Derekas}, {Devor}, {Fischer},
  {Green}, {Gropp}, {Jacobs}, {Johnston}, {LaCourse}, {Saetre}, {Schwengeler},
  {Toczyski}, {Werner}, {Garrett}, {Gore}, {Martinez}, {Spitzer}, {Stevick},
  {Thomadis}, {Vrijmoet}, {Yenawine}, {Batalha}, \&
  {Borucki}}]{2016AJ....151...68K}
{Kirk}, B., {Conroy}, K., {Pr{\v s}a}, A., {et~al.} 2016, Astronomical Journal,
  151, 68

\bibitem[{{Kowalski} \& {Saumon}(2006)}]{2006ApJ...651L.137K}
{Kowalski}, P.~M., \& {Saumon}, D. 2006, Astrophysical Journal Letters, 651,
  L137

\bibitem[{{Kruse} \& {Agol}(2014)}]{2014Sci...344..275K}
{Kruse}, E., \& {Agol}, E. 2014, Science, 344, 275

\bibitem[{{Landsman} {et~al.}(1993){Landsman}, {Simon}, \&
  {Bergeron}}]{1993PASP..105..841L}
{Landsman}, W., {Simon}, T., \& {Bergeron}, P. 1993, Publications of the
  Astronomical Society of the Pacific, 105, 841

\bibitem[{{Latham}(2007)}]{2007HiA....14..444L}
{Latham}, D.~W. 2007, Highlights of Astronomy, 14, 444

\bibitem[{{Leibovitz} \& {Hube}(1971)}]{1971AaA....15..251L}
{Leibovitz}, C., \& {Hube}, D.~P. 1971, \aap, 15, 251

\bibitem[{{Lin} {et~al.}(2011){Lin}, {Rappaport}, {Podsiadlowski}, {Nelson},
  {Paxton}, \& {Todorov}}]{2011ApJ...732...70L}
{Lin}, J., {Rappaport}, S., {Podsiadlowski}, P., {et~al.} 2011, Astrophysical
  Journal, 732, 70

\bibitem[{{Maeder}(1973)}]{1973AaA....26..215M}
{Maeder}, A. 1973, Astronomy and Astrophysics, 26, 215

\bibitem[{{Markwardt}(2009)}]{2009ASPC..411..251M}
{Markwardt}, C.~B. 2009, in Astronomical Society of the Pacific Conference
  Series, Vol. 411, Astronomical Data Analysis Software and Systems XVIII, ed.
  D.~A. {Bohlender}, D.~{Durand}, \& P.~{Dowler}, 251

\bibitem[{{Marsh}(2001)}]{2001MNRAS.324..547M}
{Marsh}, T.~R. 2001, Monthly Notices of the Royal Astronomical Society, 324,
  547

\bibitem[{{Masuda}(2015)}]{2015ApJ...805...28M}
{Masuda}, K. 2015, Astrophysical Journal, 805, 28

\bibitem[{{Mathieu} \& {Geller}(2009)}]{2009Natur.462.1032M}
{Mathieu}, R.~D., \& {Geller}, A.~M. 2009, Nature, 462, 1032

\bibitem[{{Mathur} {et~al.}(2016{\natexlab{a}}){Mathur}, {Garc{\'{\i}}a},
  {Huber}, {Regulo}, {Stello}, {Beck}, {Houmani}, \& {Salabert}}]{Mathur2016}
{Mathur}, S., {Garc{\'{\i}}a}, R.~A., {Huber}, D., {et~al.} 2016{\natexlab{a}},
  Astrophysical Journal, 827, 50

\bibitem[{{Mathur} {et~al.}(2016{\natexlab{b}}){Mathur}, {Huber}, {Batalha},
  {Ciardi}, {Bastien}, {Bieryla}, {Buchhave}, {Cochran}, {Endl}, {Esquerdo},
  {Furlan}, {Howard}, {Howell}, {Isaacson}, {Latham}, {MacQueen}, \&
  {Silva}}]{2016arXiv160904128M}
{Mathur}, S., {Huber}, D., {Batalha}, N.~M., {et~al.} 2016{\natexlab{b}}, ArXiv
  e-prints, arXiv:1609.04128

\bibitem[{{Matson} {et~al.}(2015){Matson}, {Gies}, {Guo}, {Quinn}, {Buchhave},
  {Latham}, {Howell}, \& {Rowe}}]{2015ApJ...806..155M}
{Matson}, R.~A., {Gies}, D.~R., {Guo}, Z., {et~al.} 2015, Astrophysical
  Journal, 806, 155

\bibitem[{{Maxted} {et~al.}(2011){Maxted}, {Anderson}, {Burleigh}, {Collier
  Cameron}, {Heber}, {G{\"a}nsicke}, {Geier}, {Kupfer}, {Marsh}, {Nelemans},
  {O'Toole}, {{\O}stensen}, {Smalley}, \& {West}}]{2011MNRAS.418.1156M}
{Maxted}, P.~F.~L., {Anderson}, D.~R., {Burleigh}, M.~R., {et~al.} 2011,
  Monthly Notices of the Royal Astronomical Society, 418, 1156

\bibitem[{{Maxted} {et~al.}(2013){Maxted}, {Serenelli}, {Miglio}, {Marsh},
  {Heber}, {Dhillon}, {Littlefair}, {Copperwheat}, {Smalley}, {Breedt}, \&
  {Schaffenroth}}]{2013Natur.498..463M}
{Maxted}, P.~F.~L., {Serenelli}, A.~M., {Miglio}, A., {et~al.} 2013, Nature,
  498, 463

\bibitem[{{Moe} \& {Di Stefano}(2017)}]{2017ApJS..230...15M}
{Moe}, M., \& {Di Stefano}, R. 2017, \apjs, 230, 15

\bibitem[{{Muirhead} {et~al.}(2013){Muirhead}, {Vanderburg}, {Shporer},
  {Becker}, {Swift}, {Lloyd}, {Fuller}, {Zhao}, {Hinkley}, {Pineda}, {Bottom},
  {Howard}, {von Braun}, {Boyajian}, {Law}, {Baranec}, {Riddle}, {Ramaprakash},
  {Tendulkar}, {Bui}, {Burse}, {Chordia}, {Das}, {Dekany}, {Punnadi}, \&
  {Johnson}}]{2013ApJ...767..111M}
{Muirhead}, P.~S., {Vanderburg}, A., {Shporer}, A., {et~al.} 2013,
  Astrophysical Journal, 767, 111

\bibitem[{{Nauenberg}(1972)}]{1972ApJ...175..417N}
{Nauenberg}, M. 1972, Astrophysical Journal, 175, 417

\bibitem[{{Ogilvie}(2014)}]{2014ARA&A..52..171O}
{Ogilvie}, G.~I. 2014, \araa, 52, 171

\bibitem[{{Paczynski}(1976)}]{1976IAUS...73...75P}
{Paczynski}, B. 1976, in IAU Symposium, Vol.~73, Structure and Evolution of
  Close Binary Systems, ed. P.~{Eggleton}, S.~{Mitton}, \& J.~{Whelan}, 75

\bibitem[{{Parviainen}(2015)}]{2015MNRAS.450.3233P}
{Parviainen}, H. 2015, Monthly Notices of the Royal Astronomical Society, 450,
  3233

\bibitem[{{Pavlovskii} \& {Ivanova}(2015)}]{2015MNRAS.449.4415P}
{Pavlovskii}, K., \& {Ivanova}, N. 2015, Monthly Notices of the Royal
  Astronomical Society, 449, 4415

\bibitem[{{Perets}(2015)}]{2015ebss.book..251P}
{Perets}, H.~B. 2015, {The Multiple Origin of Blue Straggler Stars: Theory vs.
  Observations}, ed. H.~M.~J. {Boffin}, G.~{Carraro}, \& G.~{Beccari}, 251

\bibitem[{{Planck Collaboration} {et~al.}(2016){Planck Collaboration}, {Ade},
  {Aghanim}, {Arnaud}, {Ashdown}, {Aumont}, {Baccigalupi}, {Banday},
  {Barreiro}, {Bartlett}, \& et~al.}]{2016AaA...594A..13P}
{Planck Collaboration}, {Ade}, P.~A.~R., {Aghanim}, N., {et~al.} 2016,
  Astronomy and Astrophysics, 594, A13

\bibitem[{{Podsiadlowski}(2014)}]{2014apa..book.....G}
{Podsiadlowski}, P. 2014, {The Evolution of Binary Systems in Accretion
  Processes in Astrophysics}

\bibitem[{{Preston} \& {Sneden}(2000)}]{2000AJ....120.1014P}
{Preston}, G.~W., \& {Sneden}, C. 2000, Astronomical Journal, 120, 1014

\bibitem[{{Rahvar} {et~al.}(2011){Rahvar}, {Mehrabi}, \&
  {Dominik}}]{2011MNRAS.410..912R}
{Rahvar}, S., {Mehrabi}, A., \& {Dominik}, M. 2011, Monthly Notices of the
  Royal Astronomical Society, 410, 912

\bibitem[{{Rappaport} {et~al.}(2015){Rappaport}, {Nelson}, {Levine},
  {Sanchis-Ojeda}, {Gandolfi}, {Nowak}, {Palle}, \&
  {Prsa}}]{2015ApJ...803...82R}
{Rappaport}, S., {Nelson}, L., {Levine}, A., {et~al.} 2015, Astrophysical
  Journal, 803, 82

\bibitem[{{Rappaport} {et~al.}(1995){Rappaport}, {Podsiadlowski}, {Joss}, {Di
  Stefano}, \& {Han}}]{1995MNRAS.273..731R}
{Rappaport}, S., {Podsiadlowski}, P., {Joss}, P.~C., {Di Stefano}, R., \&
  {Han}, Z. 1995, Monthly Notices of the Royal Astronomical Society, 273, 731

\bibitem[{{Rebassa-Mansergas} {et~al.}(2007){Rebassa-Mansergas},
  {G{\"a}nsicke}, {Rodr{\'{\i}}guez-Gil}, {Schreiber}, \&
  {Koester}}]{2007MNRAS.382.1377R}
{Rebassa-Mansergas}, A., {G{\"a}nsicke}, B.~T., {Rodr{\'{\i}}guez-Gil}, P.,
  {Schreiber}, M.~R., \& {Koester}, D. 2007, Monthly Notices of the Royal
  Astronomical Society, 382, 1377

\bibitem[{{Rebassa-Mansergas} {et~al.}(2010){Rebassa-Mansergas},
  {G{\"a}nsicke}, {Schreiber}, {Koester}, \&
  {Rodr{\'{\i}}guez-Gil}}]{2010MNRAS.402..620R}
{Rebassa-Mansergas}, A., {G{\"a}nsicke}, B.~T., {Schreiber}, M.~R., {Koester},
  D., \& {Rodr{\'{\i}}guez-Gil}, P. 2010, Monthly Notices of the Royal
  Astronomical Society, 402, 620

\bibitem[{{Rebassa-Mansergas} {et~al.}(2012){Rebassa-Mansergas}, {Nebot
  G{\'o}mez-Mor{\'a}n}, {Schreiber}, {G{\"a}nsicke}, {Schwope}, {Gallardo}, \&
  {Koester}}]{2012MNRAS.419..806R}
{Rebassa-Mansergas}, A., {Nebot G{\'o}mez-Mor{\'a}n}, A., {Schreiber}, M.~R.,
  {et~al.} 2012, Monthly Notices of the Royal Astronomical Society, 419, 806

\bibitem[{{Rebassa-Mansergas} {et~al.}(2016){Rebassa-Mansergas}, {Ren},
  {Parsons}, {G{\"a}nsicke}, {Schreiber}, {Garc{\'{\i}}a-Berro}, {Liu}, \&
  {Koester}}]{2016MNRAS.458.3808R}
{Rebassa-Mansergas}, A., {Ren}, J.~J., {Parsons}, S.~G., {et~al.} 2016, Monthly
  Notices of the Royal Astronomical Society, 458, 3808

\bibitem[{{Ren} {et~al.}(2013){Ren}, {Luo}, {Li}, {Wei}, {Zhao}, {Zhao},
  {Song}, \& {Zhao}}]{2013AJ....146...82R}
{Ren}, J., {Luo}, A., {Li}, Y., {et~al.} 2013, Astronomical Journal, 146, 82

\bibitem[{{Ren} {et~al.}(2014){Ren}, {Rebassa-Mansergas}, {Luo}, {Zhao},
  {Xiang}, {Liu}, {Zhao}, {Jin}, \& {Zhang}}]{2014AaA...570A.107R}
{Ren}, J.~J., {Rebassa-Mansergas}, A., {Luo}, A.~L., {et~al.} 2014, Astronomy
  and Astrophysics, 570, A107

\bibitem[{{Rowe} {et~al.}(2010){Rowe}, {Borucki}, {Koch}, {Howell}, {Basri},
  {Batalha}, {Brown}, {Caldwell}, {Cochran}, {Dunham}, {Dupree}, {Fortney},
  {Gautier}, {Gilliland}, {Jenkins}, {Latham}, {Lissauer}, {Marcy}, {Monet},
  {Sasselov}, \& {Welsh}}]{2010ApJ...713L.150R}
{Rowe}, J.~F., {Borucki}, W.~J., {Koch}, D., {et~al.} 2010, Astrophysical
  Journal Letters, 713, L150

\bibitem[{{Sahu} \& {Gilliland}(2003)}]{2003ApJ...584.1042S}
{Sahu}, K.~C., \& {Gilliland}, R.~L. 2003, Astrophysical Journal, 584, 1042

\bibitem[{{Sandage}(1953)}]{1953AJ.....58...61S}
{Sandage}, A.~R. 1953, Astronomical Journal, 58, 61

\bibitem[{{Schmitt} {et~al.}(2014){Schmitt}, {Wang}, {Fischer}, {Jek},
  {Moriarty}, {Boyajian}, {Schwamb}, {Lintott}, {Lynn}, {Smith}, {Parrish},
  {Schawinski}, {Simpson}, {LaCourse}, {Omohundro}, {Winarski}, {Goodman},
  {Jebson}, {Schwengeler}, {Paterson}, {Sejpka}, {Terentev}, {Jacobs},
  {Alsaadi}, {Bailey}, {Ginman}, {Granado}, {Vonstad Guttormsen}, {Mallia},
  {Papillon}, {Rossi}, \& {Socolovsky}}]{2014AJ....148...28S}
{Schmitt}, J.~R., {Wang}, J., {Fischer}, D.~A., {et~al.} 2014, Astronomical
  Journal, 148, 28

\bibitem[{{Still} \& {Barclay}(2012)}]{2012ascl.soft08004S}
{Still}, M., \& {Barclay}, T. 2012, {PyKE: Reduction and analysis of Kepler
  Simple Aperture Photometry data}, Astrophysics Source Code Library, , ,
  ascl:1208.004

\bibitem[{{Thompson} {et~al.}(2016){Thompson}, {Fraquelli}, {Van Cleve}, \&
  {Caldwell}}]{2016ksci.rept....9T}
{Thompson}, S.~E., {Fraquelli}, D., {Van Cleve}, J.~E., \& {Caldwell}, D.~A.
  2016, {Kepler Archive Manual}, Tech. rep.

\bibitem[{{Tremblay} {et~al.}(2011){Tremblay}, {Bergeron}, \&
  {Gianninas}}]{2011ApJ...730..128T}
{Tremblay}, P.-E., {Bergeron}, P., \& {Gianninas}, A. 2011, Astrophysical
  Journal, 730, 128

\bibitem[{{Trimble} \& {Thorne}(1969)}]{1969ApJ...156.1013T}
{Trimble}, V.~L., \& {Thorne}, K.~S. 1969, Astrophysical Journal, 156, 1013

\bibitem[{{Uehara} {et~al.}(2016){Uehara}, {Kawahara}, {Masuda}, {Yamada}, \&
  {Aizawa}}]{2016ApJ...822....2U}
{Uehara}, S., {Kawahara}, H., {Masuda}, K., {Yamada}, S., \& {Aizawa}, M. 2016,
  Astrophysical Journal, 822, 2

\bibitem[{{van Kerkwijk} {et~al.}(2010){van Kerkwijk}, {Rappaport}, {Breton},
  {Justham}, {Podsiadlowski}, \& {Han}}]{2010ApJ...715...51V}
{van Kerkwijk}, M.~H., {Rappaport}, S.~A., {Breton}, R.~P., {et~al.} 2010,
  Astrophysical Journal, 715, 51

\bibitem[{{Verbunt} \& {Rappaport}(1988)}]{1988ApJ...332..193V}
{Verbunt}, F., \& {Rappaport}, S. 1988, Astrophysical Journal, 332, 193

\bibitem[{{Wang} {et~al.}(2015){Wang}, {Fischer}, {Barclay}, {Picard}, {Ma},
  {Bowler}, {Schmitt}, {Boyajian}, {Jek}, {LaCourse}, {Baranec}, {Riddle},
  {Law}, {Lintott}, {Schawinski}, {Simister}, {Gr{\'e}goire}, {Babin}, {Poile},
  {Jacobs}, {Jebson}, {Omohundro}, {Schwengeler}, {Sejpka}, {Terentev},
  {Gagliano}, {Paakkonen}, {Otnes Berge}, {Winarski}, {Green}, {Schmitt},
  {Kristiansen}, \& {Hoekstra}}]{2015ApJ...815..127W}
{Wang}, J., {Fischer}, D.~A., {Barclay}, T., {et~al.} 2015, Astrophysical
  Journal, 815, 127

\bibitem[{{Welsh} {et~al.}(2011){Welsh}, {Orosz}, {Aerts}, {Brown},
  {Brugamyer}, {Cochran}, {Gilliland}, {Guzik}, {Kurtz}, {Latham}, {Marcy},
  {Quinn}, {Zima}, {Allen}, {Batalha}, {Bryson}, {Buchhave}, {Caldwell},
  {Gautier}, {Howell}, {Kinemuchi}, {Ibrahim}, {Isaacson}, {Jenkins}, {Prsa},
  {Still}, {Street}, {Wohler}, {Koch}, \& {Borucki}}]{2011ApJS..197....4W}
{Welsh}, W.~F., {Orosz}, J.~A., {Aerts}, C., {et~al.} 2011, Astrophysical
  Journals, 197, 4

\bibitem[{{Zahn}(2008)}]{2008EAS....29...67Z}
{Zahn}, J.-P. 2008, in EAS Publications Series, Vol.~29, EAS Publications
  Series, ed. M.-J. {Goupil} \& J.-P. {Zahn}, 67--90

\bibitem[{{Zorotovic} \& {Schreiber}(2013)}]{2013AaA...549A..95Z}
{Zorotovic}, M., \& {Schreiber}, M.~R. 2013, Astronomy and Astrophysics, 549,
  A95

\bibitem[{{Zorotovic} {et~al.}(2014){Zorotovic}, {Schreiber}, \&
  {Parsons}}]{2014AaA...568L...9Z}
{Zorotovic}, M., {Schreiber}, M.~R., \& {Parsons}, S.~G. 2014, Astronomy and
  Astrophysics, 568, L9

\end{thebibliography}


\listofchanges

\end{document}